\documentclass[conference]{IEEEtran}
\usepackage{amsmath,amsfonts}
\usepackage{array}
\usepackage[caption=false,font=normalsize,labelfont=sf,textfont=sf]{subfig}
\usepackage[hidelinks]{hyperref}
\usepackage{textcomp}
\usepackage{stfloats}
\usepackage{url}
\usepackage{cite}
\usepackage{flushend}
\usepackage{verbatim}
\usepackage{graphicx}
\usepackage{diagbox}
\usepackage{graphicx}
\usepackage{textcomp}
\usepackage{booktabs}
\usepackage{multirow}
\usepackage{graphicx}
\usepackage[normalem]{ulem}
\useunder{\uline}{\ul}{}
\usepackage{microtype}
\usepackage{paralist}
\usepackage{lscape}
\usepackage[table]{xcolor}
\usepackage{threeparttable} 

\newtheorem{definition}{Definition}

\usepackage{rotating}

\usepackage{array} 
\usepackage{booktabs} 
\usepackage{algorithm}
\usepackage{algpseudocode}
\algnewcommand\algorithmicforeach{\textbf{for each}}
\algdef{SE}[FUNCTION]{Function}{EndFunction}[2]{\algorithmicfunction\ \textnormal{#1}\ (#2)}{\algorithmicend\ \algorithmicfunction}%
\algdef{S}[FOR]{ForEach}[1]{\algorithmicforeach\ #1\ \algorithmicdo}
\algrenewcommand\alglinenumber[1]{\footnotesize #1}
\algrenewcommand\algorithmicrequire{\small \textbf{input:}}
\algrenewcommand\algorithmicensure{\small \textbf{output:}}
\algrenewcommand\algorithmicfunction{\textbf{Function}}
\algtext*{EndFunction} %
\algtext*{EndIf} %
\algtext*{EndFor} %
\algtext*{EndWhile} %

\usepackage{fontawesome}

\usepackage{xcolor}
\usepackage{framed}
\usepackage{textcomp}
 
\usepackage{microtype}

\usepackage{tcolorbox}
\usepackage{ifthen}
\usepackage{graphicx}
\usepackage{tcolorbox}
\usepackage{booktabs}

\newtcolorbox{custombox}[1]{
	colback=gray!10,
	colframe=gray!70,
	left=1.5mm,
	right=1.5mm,
	top=1.5mm,
	bottom=1.5mm,
	fonttitle=\bfseries,
	arc=0mm,
	leftrule=1mm,
	rightrule=0mm,
	toprule=0mm,
	bottomrule=0mm,
	notitle,
	before=\par\medskip\medskip\noindent,
	before upper={\textbf{#1: } },
}

\usepackage{xspace}
\usepackage{pifont}
\usepackage{fontawesome}
\usepackage{tikz}
\usepackage{pgfplots}
\usetikzlibrary{intersections}
\usepgfplotslibrary{fillbetween}
\pgfplotsset{compat=1.12}
\usepackage{multirow}
\usepackage{xspace}
\newcommand{\numtests}{561,267\@\xspace}
\newcommand{\numtestsunique}{178,180\@\xspace}
\newcommand{\nummr}{36\@\xspace}
\newcommand{\nummrtotal}{191\@\xspace}
\newcommand{\numpapers}{44\@\xspace}
\newcommand{\tool}{\textsc{LLMorph}\@\xspace}
\newcommand{\keyword}[1]{``\texttt{#1}''\@\xspace}
\newcommand{\mypar}[1]{\medskip\noindent \hbox{\textbf{#1}}}

\newcommand{\gpt}{\textsc{GPT-4}\@\xspace}
\newcommand{\llama}{\textsc{Llama3}\@\xspace}

\newcommand{\hermes}{\textsc{Hermes 2}\@\xspace}
\usepackage{footmisc} 

\definecolor{mygreen}{HTML}{02818a}
\definecolor{darkorange}{HTML}{FE6741}

\definecolor{nord0}{HTML}{2E3440}
\definecolor{nord1}{HTML}{3B4252}
\definecolor{nord2}{HTML}{434C5E}
\definecolor{nord3}{HTML}{4C566A}
\definecolor{nord4}{HTML}{D8DEE9}
\definecolor{nord5}{HTML}{E5E9F0}
\definecolor{nord6}{HTML}{ECEFF4}
\definecolor{nord7}{HTML}{8FBCBB}
\definecolor{nord8}{HTML}{88C0D0}
\definecolor{nord9}{HTML}{81A1C1}
\definecolor{nord10}{HTML}{5E81AC}
\definecolor{nord11}{HTML}{BF616A}
\definecolor{nord12}{HTML}{D08770}
\definecolor{nord13}{HTML}{EBCB8B}
\definecolor{nord14}{HTML}{A3BE8C}
\definecolor{nord15}{HTML}{B48EAD}

\newcommand{\stefano}[1]{\textbf{\textcolor{red}{[ \ding{46}Stefano: #1]}}}

\usepackage{letltxmacro}

\def\BibTeX{{\rm B\kern-.05em{\sc i\kern-.025em b}\kern-.08em
    T\kern-.1667em\lower.7ex\hbox{E}\kern-.125emX}}
\usepackage{balance}
\begin{document}
\title{Metamorphic Testing of Large Language Models \\ for Natural Language Processing}

\author{
    \IEEEauthorblockN{Steven Cho}
    \IEEEauthorblockA{
        University of Auckland\\
        Auckland, New Zealand\\
        steven.cho@auckland.ac.nz
    }
    \and
    \IEEEauthorblockN{Stefano Ruberto}
    \IEEEauthorblockA{
        JRC European Commission\\
        Ispra, Italy\\
        stefano.ruberto@ec.europa.eu
    }
    \and
    \IEEEauthorblockN{Valerio Terragni}
    \IEEEauthorblockA{
        University of Auckland\\
        Auckland, New Zealand\\
        v.terragni@auckland.ac.nz
    }
    }

\maketitle

\begin{abstract}
Using Large Language Models (LLMs) to perform Natural Language Processing (NLP) tasks has been becoming increasingly pervasive in recent times. The versatile nature of LLMs makes them applicable to a wide range of such tasks. While the performance of recent LLMs is generally outstanding, several studies have shown that LLMs can often produce incorrect results. Automatically identifying these faulty behaviors is extremely useful for improving the effectiveness of LLMs.

One obstacle to this is the limited availability of labeled datasets, necessitating an oracle to determine the correctness of LLM behaviors. Metamorphic Testing (MT) is a popular testing approach that alleviates this oracle problem. At the core of MT are Metamorphic Relations (MRs), defining the relationship between the outputs of 
related inputs. MT can expose faulty behaviors without the need for explicit oracles (e.g., labeled datasets).

This paper presents the most comprehensive study of MT for LLMs to date. We conducted a literature review and collected \nummrtotal MRs for NLP tasks. We implemented a representative subset (\nummr MRs) to conduct a series of experiments with three popular LLMs, running $\sim$560K metamorphic tests. The results shed light on the capabilities and opportunities of MT for LLMs, as well as its limitations.
\end{abstract}


\section{Introduction}

Large Language Models (LLMs) have rapidly gained traction in many applications due to their impressive natural-language understanding and generation capabilities~\cite{zhao2023survey}. However, concerns on the reliability and trustworthiness of LLMs persist—specifically, biases, hallucinations, and other faulty behaviors remain significant challenges~\cite{wang2024earth,guo2022threats}. Exposing such issues is crucial as it is the first step for fixing them~\cite{elazar2021measuring,chang2024survey}.

\smallskip
\textbf{Automated testing} is crucial for evaluating the quality of LLM outputs~\cite{wang2024earth,elazar2021measuring,chang2024survey,guo2022threats}. It also plays a key role in the feedback loop that continually improves LLMs over time. When testing uncovers unwanted behaviors, targeted interventions---such as reinforcement learning~\cite{sun2024inverse}, alignment adjustments~\cite{wang2023aligning}, policy updates~\cite{tang2023policygpt}, focused fine-tuning~\cite{vm2024fine}, and other adaptive techniques---can be applied to substantially enhance a model’s reliability~\cite{chang2024survey}.
Moreover, automated testing of LLMs is becoming increasingly important in industry contexts, where many organizations are transitioning away from generic, public (and supposedly well-tested) LLM services (e.g., OpenAI API) toward private, locally deployed, fine-tuned models~\cite{irugalbandara2024scaling,hanke2024open}. This shift is often driven by privacy, security, and compliance requirements, or simply by the need for better performance in company-specific environments~\cite{hanke2024open}. By rigorously testing these locally hosted models, organizations can ensure they meet performance standards while avoiding exposure of sensitive data to third-party services.

\smallskip
One of the primary challenges in automatically testing LLMs is the \textbf{oracle problem}~\cite{barr2014oracle}; \textit{the problem of distinguishing between correct and incorrect test executions}. In \textbf{Natural Language Processing (NLP)}, obtaining new text inputs for testing is straightforward thanks to abundant textual data. However, determining output correctness is far more complex. Typically, human-created labels serve as the “oracles”, indicating correct or incorrect outputs. Because human-labeled datasets for NLP are limited and expensive to obtain, there is a pressing need for effective automated test oracles~\cite{guo2022threats} that can evaluate correctness of textual outputs without relying on human-created labels.

\smallskip
\textbf{Metamorphic Testing (MT)}~\cite{1998-chen-tr} is a widely used approach to alleviate the oracle problem. At the core of MT are \textbf{Metamorphic Relations (MRs)}, which define relationships between the outputs of 
related inputs. This approach is based on the intuition that, even if we cannot automatically determine the correctness of the output for an individual input, \textit{we can use the relationships among the expected outputs of multiple related inputs as a test oracle}~\cite{2017-chen-cs}. Metamorphic testing has been used extensively in NLP, across a wide range of tasks~\cite{mr-backtrans,mr-chatbots1,mr-checklist,mr-sa,mr-mtqas}. 
However, applying MT to LLMs is still an understudied problem~\cite{mr-metal} despite its several potential benefits.


\begin{figure*}[th!]
    \centering
    \includegraphics[width=1\linewidth]{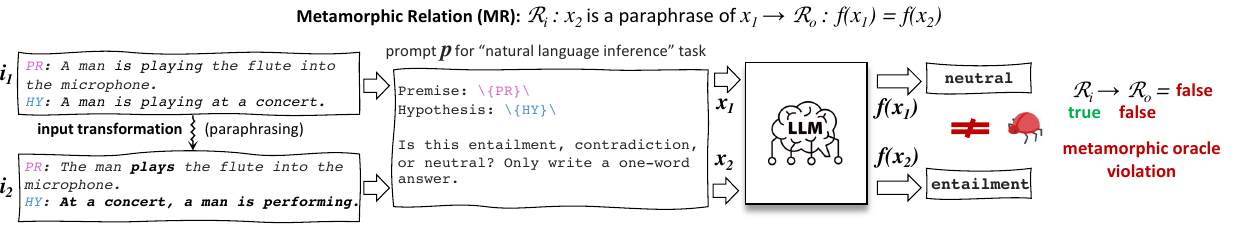}
    \vspace{-8mm}
    \caption{Real faulty behaviour detected by \tool in the OpenAI \texttt{gpt-4-1106} model.  }
    \label{fig:example}
\end{figure*}


\smallskip
Figure~\ref{fig:example} shows an example of faulty behavior we detected in \textbf{GPT-4} using MT. The example involves \textit{natural language inference}~\cite{Bowman2015}, a core NLP task that determines the relationship between a premise and a hypothesis. It classifies it as either \texttt{entailment} (hypothesis logically follows from the premise), \texttt{contradiction} (hypothesis is logically incompatible with the premise), or \texttt{neutral} (no clear logical relation). 
The MR in Figure~\ref{fig:example} states that, \textit{given two inputs where one is a paraphrase of the other, the inference results should be the same}. We create this input pair by applying a transformation on the first input to generate the second. In this case, the LLM responds \texttt{neutral} for the first input and \texttt{entailment} for the second, violating the metamorphic relation\footnote{
To note, the LLM's \texttt{neutral} response is correct, as the premise lacks evidence that the man is performing. He could be playing at home, for instance}. Notably, we can identify the faulty behavior without needing to know the expected results for either input. This highlights the key advantage of MT: the same MRs can be applied to arbitrary inputs\footnote{Inputs that satisfy the input relation, $\mathcal{R}_i$. See Section~II.}, enabling automated testing with a vast amount of data. This is crucial for testing LLMs, as faulty behavior often manifests only for specific inputs~\cite{chang2024survey}.

\smallskip
Hyun et al.~\cite{mr-metal} recently presented the only work on MT for LLMs on NLP tasks, showing its potential. They presented the framework METAL implementing 13 MRs and evaluated on six tasks on 3 LLMs.
However, questions remain about the effectiveness of MT for LLMs, as well as their true positive rate and fault detection compared to traditional ML testing. Further research is needed to fully understand the effectiveness and limitations of MT for LLMs.

\smallskip
To address these gaps, this paper presents the most comprehensive study on MT for LLMs. We conducted the first systematic literature search on MRs for NLP, reviewing 1,024 papers and identifying \numpapers that explicitly defined MRs. This resulted in a catalog of \nummrtotal unique MRs across 24 tasks. We then developed \textbf{\tool}, an automated framework for applying MT to LLMs, implementing a representative subset of \nummr MRs. Finally, we conducted large-scale experiments using \tool on three LLMs (\gpt, \llama, and \hermes) and four datasets, leading to \numtests test executions. Our experiments yielded four key findings.

\smallskip
First, MT effectively exposes faulty behaviors, with an average failure rate of 18\%. 
While a ground-truth oracle detects more faults, it requires costly manual labeling. MT, however, works on unlabeled data and identifies 11\% of failures missed by traditional testing, highlighting its complementarity.


\smallskip
Second, a manual analysis of 937 metamorphic oracle violations showed an average true positive (FP) rate of around 60\%, with most false positives arising from the intrinsic limitations of MT for NLP. 
The FP rate aligns with traditional MT for NLP, showing no LLM-specific issues, and remains unavoidable even with LLM-based input transformations.


\smallskip
Third, MR effectiveness depends on the relation and task, with minimal variation across LLMs. While the FP rate can be high, many MRs maintain an acceptable FP rate when applied to LLMs, making them practical for testing. In particular,  some MRs are consistently more effective than others, with high failure rate and low false positive rate (e.g., MRs 9, 142, 154). Developers could prioritise these during testing. 

\smallskip
Finally, our results show that
several MRs are task-independent and thus useful for evaluating fine-tuned LLMs. Additionally, LLM flakiness is not a major concern.


\smallskip
In summary, this paper makes the following contributions:

\begin{itemize}
    \item An exhaustive catalog of \nummrtotal MRs for NLP tasks\footnote{\label{link:catalog}\url{https://mt4nlp.github.io/}}.
    \item \tool, a framework for performing MT on LLMs, implementing \nummr MRs, of which can be easily extended. 
    \item A series of experiments that shed light on the capabilities and limitations of MT for LLMs.
    \item Release of the source code for \tool\footnote{\label{link:gh}\url{https://github.com/steven-b-cho/llmorph}} and all experimental data\footnote{\label{link:doi}\url{https://doi.org/10.5281/zenodo.16526643}} to foster future work in this area

\end{itemize}

\section{Metamorphic Testing for LLMs}

This section provides the background of this work and formulates the problem of MT for LLMs. We follow the traditional definition of MRs by Chen et al.~\cite{1998-chen-tr,2017-chen-cs}:

\medskip
\begin{definition}
Let $f$ be a target algorithm. A \textbf{Metamorphic Relation (MR)} is a property of $f$ involving multiple inputs $\langle x_1, \dots, x_n \rangle$ and their outputs $\langle f(x_1), \dots, f(x_n) \rangle$, with $n \geq 2$. Formally\footnote{A more general definition of MRs allows inputs and outputs to also appear in the output and input relations, respectively~\cite{2017-chen-cs}. However, we use the traditional definition (Definition 1) as it covers almost all collected MRs. Exceptions are MRs 115-119, 165-166.}, an MR is expressed as a logical implication 
$\mathcal{R}_i \Rightarrow \mathcal{R}_o$ 
from an \textbf{input relation} $\mathcal{R}_i$ to an \textbf{output relation} $\mathcal{R}_o$
\vspace{-1mm}
\[ 
\mathcal{R}_i \left( x_1, \cdots x_n \right) \Rightarrow \mathcal{R}_o(f(x_1), \cdots f(x_n)) 
\]
\end{definition}

\smallskip
Whenever a given input relation $\mathcal{R}_i$ holds between two or more inputs, a corresponding output relation $\mathcal{R}_o$ is expected to hold between the outputs. If $\mathcal{R}_i$ is true, then $\mathcal{R}_o$ should be true. Thus, an MR can be used as a metamorphic test oracle.

\medskip
\begin{definition}
Given an MR ($\mathcal{R}_i \Rightarrow \mathcal{R}_o$), a \textbf{metamorphic test oracle} is an executable Boolean expression that reports a faulty behavior if, for a specific set of inputs, the input relation $\mathcal{R}_i$ is satisfied (true), but the output relation $\mathcal{R}_o$ is not (false).
\end{definition}

\smallskip
For example, the MR of Figure~\ref{fig:example} is formulated as: $\mathcal{R}_i := x_2 \textit{ is a paraphrase of } x_1\Rightarrow\mathcal{R}_o := f(x_1) = f(x_2)$. It reports a faulty behaviour in Figure~\ref{fig:example} because the input relation is satisfied while the output relation is not.

\smallskip
Typically, metamorphic test cases are generated by first obtaining an initial test input $x_1$ (existing or generated), then applying a transformation~\cite{2018-zhou-tse}~$\rightsquigarrow$ to $x_1$ to create new inputs that satisfy the input relation~\cite{xu-tosem-2024}. In such cases, the initial test input is called the \textbf{source test input} (or case), which is $x_1$ in Figure~\ref{fig:example}. The case(s) derived from it to satisfy the input relation are called the \textbf{follow-up test inputs} (or cases), which is $x_2$ in the example.  We can thus formalise this as:

\medskip
\begin{definition}
Given an MR ($\mathcal{R}_i \Rightarrow \mathcal{R}_o$) and a test input $x_1$, a \textbf{(metamorphic) input transformation}~\cite{ayerdi-genmorph-tse-2024} $\mathbf{\rightsquigarrow}$  is a transformation $x_1 \rightsquigarrow x_2$ that satisfies $R_i$ (i.e., $x_1 \rightsquigarrow x_2 : \mathcal{R}_i(x_1, x_2) = \text{true}$) 
\end{definition}


\medskip
Given a single MR, MT can consider an arbitrary number of source test inputs and use the \emph{input transformation} to automatically create the corresponding follow-up test inputs. MT executes the function under test on each pair of inputs and reports an oracle violation if the output relation is false.

\smallskip
To apply \textbf{MT for LLMs}, we need a few considerations. First, we consider only the use of LLMs for NLP---that is, both the input $x$ and the output $f(x)$ are natural language text. Second, as our target algorithm \textit{f} is an LLM, it only provides next token prediction. To perform a task, LLMs require specific \textbf{prompts} (instructions). In the example in Figure~\ref{fig:example}, the prompt~$p$ is given for the natural language inference task, and is part of the input $x$ to execute the task with an LLM. To obtain comparable results among outputs, the prompt must remain consistent between the source and follow-up inputs; it should not be altered by the input transformation. 
LLM prompt perturbation is a separate topic~\cite{li2023multistepjailbreakingprivacyattacks} that falls outside the scope of this work. Formally, an input $x$ is a tuple $\langle i, p \rangle$, where $i$ is the textual input and $p$ is the prompt (instructions) on how to perform the NLP task using the input $i$. The input transformation $\rightsquigarrow$ ($\mathcal{R}_i$) only modifies (predicates on) the textual input $i$ (see Figure~\ref{fig:example}).

\section{Literature Search of MRs}


\begin{table}[t]
\caption{Top publications venues of the reviewed publications}
\label{tab:venue_table}
\rowcolors{2}{gray!10}{white} 


\setlength{\tabcolsep}{4pt}
\renewcommand{\arraystretch}{1}
\resizebox{\columnwidth}{!}{%
\begin{threeparttable}

\begin{tabular}{p{0.9\linewidth} cp{0.1\linewidth}} \toprule
\textbf{Venue }                                    & \multicolumn{1}{r}{\textbf{\# Papers}} \\ \midrule
International   Conference on Software Engineering (ICSE)              & ~5   \\
International Conference on Automated Software Engineering (ASE)       & ~5   \\
International Conf. on the Foundations of Software Engineering (FSE)
& ~4   \\
arXiv                                                                  & ~4   \\
International Workshop on Metamorphic Testing (MET)                    & ~3   \\
ACM Transactions on Software Eng. and Methodology  (TOSEM)                & ~2   \\
International Computer Software and Applications Conf. (COMPSAC)  & ~2   \\
Mathematics                                                            & ~2   \\ 

Others & 17 \\ \hiderowcolors
\midrule
\textbf{Total} & \textbf{44} \\ 

\bottomrule

\end{tabular}%
\end{threeparttable}
}
\end{table}

Our objective is to test LLMs using MRs for tasks involving natural language input and output. To that end, we systematically search for literature applying MT in natural language-based systems to create a comprehensive list of such relations.


\smallskip
\noindent
\textbf{Literature search:} For our search, we follow the ACM/SIGSOFT systematic review standards\footnote{\url{https://www2.sigsoft.org/EmpiricalStandards/docs/standards?standard=SystematicReviews}} as closely as possible, though though some procedures are inapplicable as our goal was is search, not a review.
As MT for NLP and LLMs is a relatively new area, we aimed for comprehensiveness and queried Google Scholar, which indexes both formal digital libraries (e.g., ACM, IEEE)~\cite{zhang2023scholarly} and preprint servers (e.g., arXiv).  
We used the following keyword search:

\medskip
{\footnotesize 
[\keyword{metamorphic test} $\lor$ \keyword{metamorphic testing} $\lor$ \keyword{metamorphic relation(s)}] $\wedge$ [\keyword{llm(s)} $\lor$ \keyword{language model}  $\lor$ \keyword{nlp} $\lor$ \keyword{natural language processing}]
}
\medskip

We include \keyword{llm} and \keyword{language model} as we wish to also find MRs that have already been used for testing LLMs and other language models.
We searched papers indexed from 1 January 1998, the year when metamorphic testing was first introduced~\cite{1998-chen-tr}, to 30 June 2024. 
This resulted in 1,024 papers.

\smallskip
\noindent
\textbf{Literature selection:}
The literature search we performed was intentionally broad, as we intended to gather as many MRs as possible. Many of the papers chosen were thus out of scope. To filter these, we included only papers that use MT to test a natural language-based system explicitly mentioning the metamorphic relations used. We excluded papers not written in English, and those that only test for fairness or biases, which are outside our scope.


\smallskip
After this inspection, we performed backward and forward snowballing, examining references to identify any relevant papers we missed. The final result is \textbf{\numpapers papers}.

\begin{figure}[t!]
\centering
  \includegraphics[width=0.70\linewidth]{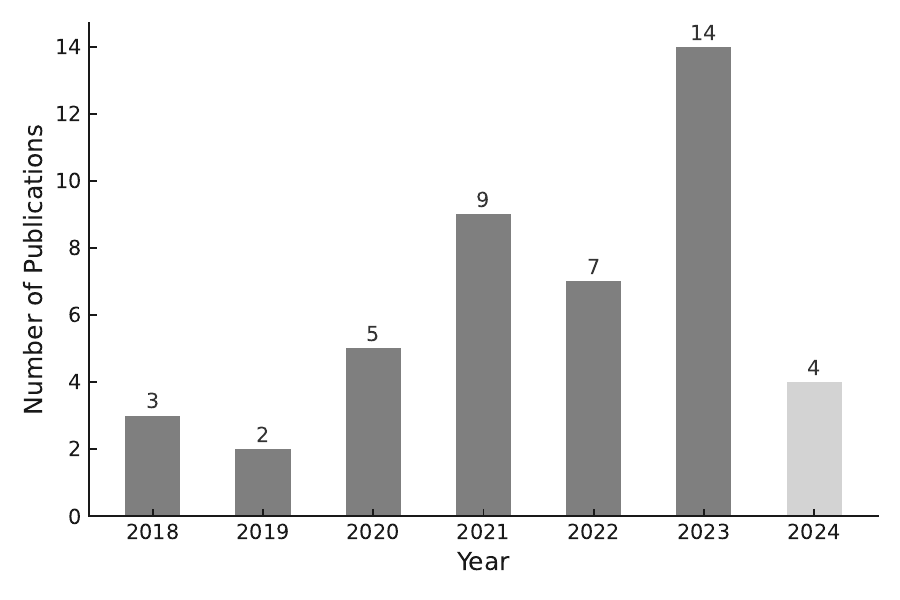}
  \vspace{-5mm}
  \caption{Number of the \numpapers papers published each year from 2018 to 2024}
  \vspace{-1.5mm}
  \label{fig:years_chart}
\end{figure}

\begin{table*}[htbp]\centering
\caption{191 Metamorphic relations for NLP tasks. 
}
\vspace{-2mm}
\label{tab:mr_table}

\rowcolors{2}{gray!10}{white} 


\setlength{\tabcolsep}{1pt}
\renewcommand{\arraystretch}{1}

\resizebox{\linewidth}{!}{%
\begin{threeparttable}
\begin{tabular}{rlllc||rlllc}
\toprule

\textbf{ID} & \textbf{Task} & \textbf{Input Relation ($\mathbf{\mathcal{R}_i}$)} & \textbf{Output Relation ($\mathbf{\mathcal{R}_o}$)} & \textbf{Ref.} & \textbf{ID} & \textbf{Task} & \textbf{Input Relation ($\mathbf{\mathcal{R}_i}$) } & \textbf{Output Relation ($\mathbf{\mathcal{R}_o}$)} & \textbf{Ref.} \\ \midrule
1 & $G_a$ & Replace characters with random & Equivalence & \cite{mr-metal} & 97 & QAm & Add negation (for multi-choice) & Difference & \cite{mr-chinesellm} \\
2 & $G_a$ & Delete characters & Equivalence & \cite{mr-metal} & 98 & QAm & Add background for testing   industrial sectors to question & Equivalence & \cite{mr-chinesellm} \\
3 & $G_a$ & Leet format conversion & Equivalence & \cite{mr-metal} & 99 & QAm & Add security instruction to   question & Equivalence & \cite{mr-chinesellm} \\
4 & $G_a$ & Add random characters & Equivalence & \cite{mr-metal} & 100 & QAm & Add irrelevant option & Equivalence & \cite{mr-chinesellm} \\
5 & $G_b$ & Add spaces & Equivalence & \cite{mr-metal} & 101 & QAc & Semantic invariance & Equivalence & \cite{mr-intergen} \\
6 & $G_b$ & Swap characters & Equivalence & \cite{mr-metal} & 102 & QAc & Capitalisation (in question) & Equivalence & \cite{mr-mtqas} \\
7 & $G_a$ & Shuffle characters & Equivalence & \cite{mr-metal} & 103 & QAc & Rephrase comparative sentence (in   question) & Equivalence & \cite{mr-mtqas} \\
8 & $G_a$ & Synonym substitution & Equivalence & \cite{mr-metal} & 104 & QAc & Replace comparative word with   antonym (in question) & Difference & \cite{mr-mtqas} \\
9 & $G_a$ & Word insertion & Equivalence & \cite{mr-metal} & 105 & QAc & Replace subject with unrelated   noun (in question) & Difference & \cite{mr-mtqas} \\
10 & $G_a$ & Antonyms substitution & Difference & \cite{mr-metal} & 106 & QAc & Capitalisation (in context) & Equivalence & \cite{mr-mtqas} \\
11 & $G_c$ & Remove sentences & Equivalence & \cite{mr-metal} & 107 & QAc & Reverse order of sentences (in   context) & Equivalence & \cite{mr-mtqas} \\
12 & $G_c$ & Replace sentences & Equivalence & \cite{mr-metal} & 108 & QAc & Add irrelevant sentences (in   context) & Equivalence & \cite{mr-mtqas} \\
13 & $G_d$ & Category-based substitution   (names, pronouns) & Equivalence & \cite{mr-chinese} & 109 & QAc & Remove irrelevant sentences (in   context) & Equivalence & \cite{mr-mtqas} \\
14 & $G_d$ & Category-based substitution   (country) & Equivalence & \cite{mr-chinese} & 110 & QAc & Category-based substitution (numbers) (in context) & Equivalence*& \cite{mr-mtqas} \\
15 & $G_d$ & Category-based substitution   (occupation) & Equivalence & \cite{mr-chinese} & 111 & QAc & Add irrelevant sentence (to   question) & Equivalence & \cite{mr-nattestgen} \\
16 & $G_d$ & Replace punctuation & Equivalence & \cite{mr-chinese} & 112 & QAc & Add irrelevant sentence (to   context) & Equivalence & \cite{mr-nattestgen} \\
17 & $G_d$ & Synonym substitution (noun) & Equivalence & \cite{mr-chinese} & 113 & QAc & Combine two questions and contexts & Answers both questions & \cite{mr-nattestgen} \\
18 & $G_d$ & Synonym substitution (verb) & Equivalence & \cite{mr-chinese} & 114 & QAc & Combine two questions and contexts & Answers first question & \cite{mr-nattestgen} \\
19 & $G_d$ & Shuffle sentences & Equivalence & \cite{mr-chinese} & 115 & QA & Given a wh-ques. and answer,   derive a wh-ques. & Consistency & \cite{mr-qaasker2} \\
20 & $G_e$ & Append full stops to words or   sentences & Equivalence & \cite{mr-noise} & 116 & QA & Given a wh-ques. and answer,   derive a gen-ques. & Consistency & \cite{mr-qaasker2} \\
21 & CM & Substitute characters with   visually similar characters & Equivalence & \cite{mr-mttm} & 117 & QA & Given a gen- or alt-question and   answer, derive a wh-ques. & Consistency & \cite{mr-qaasker2} \\
22 & CM & Split characters into visually   similar characters & Equivalence & \cite{mr-mttm} & 118 & QA & Given a wh-ques., answer   and extra info., derive a wh-ques. & Consistency & \cite{mr-qaasker2} \\
23 & CM & Combine characters into visually   similar characters & Equivalence & \cite{mr-mttm} & 119 & QA & Given a wh-ques., answer   and extra info., derive a gen-ques. & Consistency & \cite{mr-qaasker2} \\
24 & CM & Add random characters & Equivalence & \cite{mr-mttm} & 120 & QA & Back translate & Equivalence & \cite{mr-qatest} \\
25 & CM & Replace characters with special   symbols & Equivalence & \cite{mr-mttm} & 121 & QA & Change position of adverbial   clause & Equivalence & \cite{mr-qatest} \\
26 & CM & Swap characters & Equivalence & \cite{mr-mttm} & 122 & QA & Word insertion & Equivalence & \cite{mr-qatest} \\
27 & CM & Translate some words to other   languages & Equivalence & \cite{mr-mttm} & 123 & QA & Synonym substitution & Equivalence & \cite{mr-qatest} \\
28 & CM & Homophone substitution & Equivalence & \cite{mr-mttm} & 124 & QA & Alias substitution & Equivalence & \cite{mr-qatest} \\
29 & CM & Abbreviation substitution & Equivalence & \cite{mr-mttm} & 125 & QA & Abbreviation replacement & Equivalence & \cite{mr-qatest} \\
30 & CM & Split words into sub-words & Equivalence & \cite{mr-mttm} & 126 & QA & Keyboard mistake & Equivalence & \cite{mr-qatest} \\
31 & CM & Add many irrelevant sentences & Equivalence & \cite{mr-mttm} & 127 & QA & Misspelled word & Equivalence & \cite{mr-qatest} \\
32 & DS & Synonym substitution & Equivalence & \cite{mr-chatbots2} & 128 & QA & OCR recognition error & Equivalence & \cite{mr-qatest} \\
33 & DS & Category-based substitution (numbers) & Equivalence*& \cite{mr-chatbots2} & 129 & QA & Repeated question mark & Equivalence & \cite{mr-qatest} \\
34 & DS & Remove keywords & Difference & \cite{mr-chatbots2} & 130 & QAb & Antonym substitution (first   adjective) & Difference & \cite{mr-readcomp} \\
35 & DS & Replace keywords with unrelated   terms & Difference & \cite{ mr-chatbots2} & 131 & QAb & Change tense & Difference & \cite{mr-readcomp} \\
36 & DS & Change order of actions & Difference & \cite{mr-chatbots2} & 132 & QAb & Substitute between ‘before’ and   ‘after’ & Difference & \cite{mr-readcomp} \\
37 & DS & Back translate & Equivalence & \cite{mr-dialtest} & 133 & QAb & Add negation (for boolean) & Difference & \cite{mr-readcomp} \\
38 & DS & Word insertion & Equivalence & \cite{mr-dialtest} & 134 & QAb & Synonym substitution (adjective) & Equivalence & \cite{mr-readcomp} \\
39 & DSp & Synonymous sentence substitution   in persona & Equivalence & \cite{mr-persona} & 135 & QAb & Change adverbial position & Equivalence & \cite{mr-readcomp} \\
40 & DSp & Persona substitution & Equivalence & \cite{mr-persona} & 136 & QAb & Change to active/passive voice & Equivalence & \cite{mr-readcomp} \\
41 & DSp & Swap characters in only persona & Equivalence & \cite{mr-persona} & 137 & RE & Category-based substitution & Equivalence & \cite{mr-mtre} \\
42 & DSp & Swap characters in only input & Equivalence & \cite{mr-persona} & 138 & RE  & Replace head/tail with   coarser-grained entity & Same or coarser relation & \cite{mr-mtre} \\
43 & DSp & Swap characters in both persona   and input & Equivalence & \cite{mr-persona} & 139 & RE & Replace head/tail with co-related   entity of different type & Equivalence & \cite{mr-mtre} \\
44 & CR & Synonym, antonym and mask   substitution & Consistency & \cite{mr-coref} & 140 & RE & Replace head/tail with   coreferential entity & Equivalence & \cite{mr-mtre} \\
45 & FN & Back translate & Equivalence & \cite{mr-fn1} & 141 & RE & Swap entities if they have a   symmetrical relation & Equivalence & \cite{mr-mtre} \\
46 & FN & Shuffle sentences & Equivalence & \cite{mr-fn2} & 142 & RE & Swap entities if they have an   asymmetrical relation & Opposite relation & \cite{mr-mtre} \\
47 & FN & Shuffle words within a sentence & Difference in confidence & \cite{mr-fn2} & 143 & RE & $f(x_1, x_2)$ is asymmetrical and   $f(x_1, x_3)$ is symmetrical & $f(x_3, x_2)$ equals $f(x_1, x_2)$ & \cite{mr-mtre} \\
48 & FN & Add/remove negation & Difference & \cite{mr-fn2} & 144 & RE & $f(x_1, x_2)$ is asymmetrical and   $f(x_2, x_3)$ is symmetrical & $f(x_1, x_3)$ equals $f(x_1, x_2)$ & \cite{mr-mtre} \\
49 & FN & No transformation & Equivalence & \cite{mr-fn2} & 145 & SA & $x_1$ is more pos./neg. than $x_2$, add random sentence to both& $x_3$ more pos./neg. than $x_4$ & \cite{mr-sct} \\
50 & FN & Give input twice & Equivalence & \cite{mr-fn2} & 146 & SA & Add positive/negative text & More positive/negative & \cite{mr-intergen} \\
51 & FN & Paraphrase text & Equivalence & \cite{mr-fn2} & 147 & SA & Abbreviation / contraction   substitution & Equivalence & \cite{mr-sa} \\
52 & FN & Split a false $x_1$ into multiple   inputs & At least one output is false & \cite{mr-fn2} & 148 & SA & Synonym substitution (nouns) & Equivalence & \cite{mr-sa} \\
53 & FN & Combine two inputs in which at   least one is false & Combination is false & \cite{mr-fn2} & 149 & SA & Change singular/plural & Equivalence & \cite{mr-sa} \\
54 & LSR & $f(x_1, x_2)$ and $f(x_2, x_3)$   are true & $f(x_1, x_3)$ is true & \cite{mr-sct} & 150 & SA & Change ‘.’ to ‘!’ & Stronger & \cite{mr-sa} \\
55 & NER & Mask substitution (verbs,   adjectives, noun phrases) & Equivalence & \cite{mr-autoner} & 151 & SA & Add emphasising adverbs & Stronger & \cite{mr-sa} \\
56 & NER & Mask substitution (noun phrases) & Equivalence & \cite{mr-autoner} & 152 & SA & Add negation & Difference & \cite{mr-sa} \\
57 & NER & Declarative sentence into   interrogative sentence & Is a declarative sentence & \cite{mr-autoner} & 153 & SA & Reverse case of all characters & Equivalence & \cite{mr-sa} \\
58 & NER & Shuffle same-category entities in a sentence & Equivalence & \cite{mr-autoner} & 154 & SA & Make nouns and adjectives   uppercase & Stronger & \cite{mr-sa} \\
59 & NER & Transform company names to sub-companies & Recognise as sub-company & \cite{mr-mtner} & 155 & SA & Change tense & Equivalence & \cite{mr-sa} \\
60 & NER & Swap entities with the same type   but different identities & Identify the new identities & \cite{mr-mtner} & 156 & SA & Swap phrases around joining words & Equivalence & \cite{mr-sa} \\
61 & NER & Add sentence to another sentence & Union of individual inputs & \cite{mr-bioentity} & 157 & SA & Substitute between ‘although’ and   ‘but’ & Equivalence & \cite{mr-sa} \\
62 & NER & Add sentence to a paragraph & Union of individual inputs & \cite{mr-bioentity} & 158 & SA & Rephrase comparative sentence & Equivalence & \cite{mr-sa} \\
63 & NER & Add paragraph to an article & Union of individual inputs & \cite{mr-bioentity} & 159 & SA & Swap comparative objects & Difference & \cite{mr-sa} \\
64 & NER & Add list of random words to   another list of random words & Union of individual inputs & \cite{mr-bioentity} & 160 & SA & Shuffle sentences & Equivalence & \cite{mr-sa} \\
65 & NER & Remove list of words from sentence & $x_1$ minus removed output & \cite{mr-bioentity} & 161 & SA & Group sentences of same sentiment   together & Equivalence & \cite{mr-sa} \\
66 & NER & Remove sentence from paragraph & $x_1$ minus removed output & \cite{mr-bioentity} & 162 & SA & Group sentences by sentiment and   confidence & Equivalence & \cite{mr-sa} \\
67 & NER & Remove paragraph from article & $x_1$ minus removed output & \cite{mr-bioentity} & 163 & SA & Sequentially add positive   sentences & Gradually more positive & \cite{mr-sa} \\
68 & NER & Remove some words from list of   random words & $x_1$ minus removed output & \cite{mr-bioentity} & 164 & SA & Sequentially add negative   sentences & Gradually more negative & \cite{mr-sa} \\
69 & NER & Shuffle paragraphs in an article & Equivalence & \cite{mr-bioentity} & 165 & SA & Remove all sentences with same   label as document & Difference & \cite{mr-sa} \\
70 & NER & Shuffle words in a list of random   words & Equivalence & \cite{mr-bioentity} & 166 & SA & Remove all sentences with   different label than document & Equivalence & \cite{mr-sa} \\
71 & NLI & Synonym substitution (in both   premise and hypothesis) & Equivalence & \cite{mr-evalnli} & 167 & SD & $x_1$ implies $x_2$ and $x_1$   implies $x_3$ & $x_2$ implies $x_3$ & \cite{mr-stance} \\
72 & NLI & Synonym substitution (in premise) & Equivalence & \cite{mr-evalnli} & 168 & TS & Deriving an abstract sentence/summary from the  input & Difference & \cite{mr-intergen} \\
73 & NLI & Synonym substitution (in   hypothesis) & Equivalence & \cite{mr-evalnli} & 169 & TR & Translate to a different language   first & Equivalence & \cite{mr-montecarlo} \\
74 & NLI & Substitute gendered words with   words of same gender & Equivalence & \cite{mr-evalnli} & 170 & TR & Category-based substitution & Equivalence*& \cite{mr-mt4mt} \\
75 & NLI & Substitute gendered words with   words of different gender & Equivalence & \cite{mr-evalnli} & 171 & TR & Context-similar substitution & Equivalence & \cite{mr-automt} \\
76 & NLI & Add negation (in hypothesis) & Opposite & \cite{mr-evalnli} & 172 & TR & Mask substitution (nouns,   adjectives) & Structured format is similar & \cite{mr-structinv} \\
77 & NLI & Conjoin two premises that both   imply hypothesis & Equivalence & \cite{mr-evalnli} & 173 & TR & Replace words with random & Difference & \cite{mr-patinv} \\
78 & NLI & Conjoin two hypotheses that are   both implied by premises & Equivalence & \cite{mr-evalnli} & 174 & TR & Remove words & Difference & \cite{mr-patinv} \\
79 & NLI & $x_1$ implies $x_2$ and $x_2$   implies $x_3$ & $f(x_1, x_3)$ is not contr. & \cite{mr-evalnli} & 175 & TR & Reverse case of some characters & Equivalence & \cite{mr-deepl} \\
80 & NLI & $x_1$ implies $x_2$ and $x_1$   implies $x_3$ & $f(x_2, x_3)$ is not contr. & \cite{mr-evalnli} & 176 & TR & Reverse case of all characters & Equivalence & \cite{mr-googletrans} \\
81 & PST & Change position of ‘because’   clause & Equivalence & \cite{mr-pos} & 177 & TR & Remove full stops & Equivalence & \cite{mr-googletrans} \\
82 & PST & Change position of ‘when’ clause & Equivalence & \cite{mr-pos} & 178 & TR & Replace punctuation & Equivalence & \cite{mr-deepl} \\
83 & PST & Add irrelevant sentence to   beginning of input & Equivalence & \cite{mr-pos} & 179 & TR & Add noise & Equivalence & \cite{mr-googletrans} \\
84 & PST & Add irrelevant sentence to end of   input & Equivalence & \cite{mr-pos} & 180 & TR & Category-based substitution   (subject) & Equivalence*& \cite{mr-googletrans} \\
85 & PD & Paraphrase text & Equivalence & \cite{mr-plagiarism} & 181 & TR & Synonym substitution (verb) & Equivalence & \cite{mr-googletrans} \\
86 & PDq & Swap first names and surnames & Equivalence & \cite{mr-plagiarism2} & 182 & TR & Category-based substitution   (object) & Equivalence*& \cite{mr-googletrans} \\
87 & PDq & Reverse order of query & Equivalence & \cite{mr-plagiarism2} & 183 & TR & Random word substitution (word at   end of sentence) & Equivalence*& \cite{mr-deepl} \\
88 & PDq & Swap pairs of numbers & Equivalence & \cite{mr-plagiarism2} & 184 & TR & Translate to target language and   back first & Equivalence & \cite{mr-deepl} \\
89 & PDq & Abbreviate words & Equivalence & \cite{mr-plagiarism2} & 185 & TR & Place a phrase in different contexts & Phrase is equivalent & \cite{mr-reftrans} \\
90 & PDq & Category-based substitution (nationality, gender, numbers) & Difference & \cite{mr-plagiarism2} & 186 & TR & Insert adjunct sentence & Parse tree of $x_1$ found in $x_2$ & \cite{mr-constinv} \\
91 & PDq & Category-based substitution (gender, numbers) & Difference & \cite{mr-plagiarism2} & 187 & TR & Place a phrase in   different contexts then back translate & Phrase is equivalent & \cite{mr-backtrans} \\
92 & PDq & Category-based substitution (numbers) & Difference & \cite{mr-plagiarism2} & 188 & TR & Back translate & Proper nouns are the same & \cite{mr-retromorphic} \\
93 & QAm & Numeric magnitude change & Equivalence & \cite{mr-chinesellm} & 189 & TR & Construct word closures & Word closure comparison & \cite{mr-wordclosure} \\
94 & QAm & Numeric precision change & Equivalence & \cite{mr-chinesellm} & 190 & TR & Convert to tree then prune & Similarity of trees & \cite{mr-treeprune} \\
95 & QAm & Synonym substitution   (industry-specific) & Equivalence & \cite{mr-chinesellm} & 191 & TR & Back translate & Equivalence & \cite{mr-enhance} \\
96 & QAm & Swapping order of options & Equivalence & \cite{mr-chinesellm} &  &  &  &  & \\ 

\bottomrule
\end{tabular} 
\begin{tablenotes}
\footnotesize
\item[a] To improve table readability we grouped the tasks SA, TC, TD, QA, SM, IR into $G_a$, SA, TC, TD, QA into $G_b$, SM, IR into $G_c$, TS, SM, TC into $G_d$, and 
TR, NER into $G_e$
\item[b] Equivalence* means equivalence except for substitute
\item[c] Four papers: \cite{mr-chatbots2, mr-fn2, mr-qaasker2, mr-deepl}) are extensions of others: \cite{mr-chatbots1, mr-fn1, mr-qaasker1, mr-googletrans}) of the same authors and share many MRs. Therefore, these shared MRs are attributed only to the latest paper.
\item[d] If we found the same MR in multiple papers applied on the same task we cite the most recent one. 
\end{tablenotes}
\end{threeparttable}
}
\end{table*}

\begin{table*}[t]\centering
\caption{Description and breakdown of the Natural Language Processing (NLP) Tasks of the MRs in Table~\ref{tab:mr_table} }
\label{tab:task}
\rowcolors{2}{gray!10}{white} 


\setlength{\tabcolsep}{22pt}
\renewcommand{\arraystretch}{0.90}

\resizebox{\linewidth}{!}{%
\begin{tabular}{lllr}
\toprule
\textbf{ID} & \textbf{Task Name} & \textbf{Task Description} & \textbf{\# MRs} \\
\midrule
CM & Content moderation & Detecting inappropriate or harmful content in a text & 11 \\
CR & Coreference resolution & Determining which words in a text refer to the same entity & 1 \\
DS & Dialogue system & Conversing with humans & 7 \\
DSp & Dialogue system (persona-based) & Conversing with humans while maintaining a consistent   personality & 5 \\
FN & Fake news detection & Detecting whether a text has false or misleading information & 9 \\
IR & Information retrieval & Finding relevant information from a text & 10 \\
LSR & Lexical semantic relations & Analysing relationship between words & 1 \\
NER & Named entity recognition & Identifying and classifying words into categories & 17 \\
NLI & Natural language inference & Determining whether a hypothesis follows from a premise & 10 \\
PD & Plagiarism detection & Detecting instances of copied or unoriginal text & 1 \\
PDq & Plagiarism detection (query-based) & Detecting instances of copied or unoriginal text using a   specific query & 7 \\
PST & Part-of-speech-tagging & Classifying words into grammatical categories & 4 \\
QA & Question answering & Answering questions in natural language, using no context & 25 \\
QAb & Question answering (boolean) & Answering true or false questions, using no context & 7 \\
QAc & Question answering (incl. context) & Answering questions in natural language, using context & 14 \\
QAm & Question answering (multi-choice) & Answering multi-choice questions, using no context & 8 \\
RE & Relation extraction & Identifying relationships between two entities in a text & 8 \\
SA & Sentiment analysis & Determining how positive or negative a text is & 32 \\
SD & Stance detection & Determining the agreement between two texts & 1 \\
SM & Summarisation & Condensing text while retaining key information & 17 \\
TC & Text classification & Determining whether a text falls into predefined categories & 17 \\
TD & Toxicity detection & Identifying whether a text is offensive or harmful & 10 \\
TR & Translation & Converting text from one language to another & 24 \\
TS & Text similarity & Determining the similarity between two texts & 8 \\
\bottomrule
\end{tabular}
}
\end{table*}

\smallskip
\noindent
\textbf{Literature analysis:}
Table~\ref{tab:venue_table} and Figure~\ref{fig:years_chart} show the top publication venues and years of the \numpapers papers found, respectively. We did not find any papers that met our criteria published before 2018, despite MT being introduced 20 years earlier. This may indicate that the application of MT on NLP has only recently begun to be explored. 
Figure~\ref{fig:years_chart} shows an upward trend in the number of published papers over time, indicating growing interest and research activity in the field.

\smallskip
Table~\ref{tab:venue_table} shows that the most frequent publication venues are ICSE, ASE and FSE, the three top SE conferences. This is logical, given that MT is fundamentally a software testing concept. Four papers (9\%) were published on \textsc{arXiv} without a corresponding peer-reviewed conference or journal publication. 
However, given that the majority of the selected papers were published in peer-reviewed venues, we believe that the quality of the results is not compromised.



\smallskip
\noindent
\textbf{Catalog of metamorphic relations:} 
Finally, we analysed the \numpapers papers and extracted \textbf{\nummrtotal unique MRs} related to NLP. To do so, the first author read each selected paper and identified all defined MRs, ensuring understanding of the MR, including any specific notation. They then reformulated the MR using our notation (for consistency) and wrote a brief description. This process ensured that all MRs in each paper were captured.  Afterwards, to ensure consistency and correctness, all authors reviewed the collated list. We collectively discussed each MR, removed duplicates, and refined descriptions; and, when necessary, we revisited the papers together to ensure accurate interpretation and categorization until reaching a consensus.

The collected MRs are summarised in Table~\ref{tab:mr_table}, with the NLP task descriptions and breakdown in Table~\ref{tab:task}. Table~\ref{tab:mr_table} presents a unique ID assigned to each MR, the input relation ($\mathcal{R}_i$), the output relation ($\mathcal{R}_o$), the NLP task for which the MR was presented, and the publication source. Note that, for improved table readability, the input relation ($\mathcal{R}_i$) implicitly describes the relation between the source and follow-up input(s), and the output relation ($\mathcal{R}_o$) among their outputs. A formal description of the \nummrtotal MRs can be found in our supplementary data.





\smallskip
\noindent
\textbf{MR discussion:}
We see that the most common output relation is \texttt{Equivalence}. This is perhaps the simplest relation an input relation could imply, leading to its prevalence. In the context of NLP, \texttt{Equivalence} can be defined in two ways: syntactic and semantic. \textit{Syntactic equivalence} occurs when the two values have the same structure or form, while \textit{semantic equivalence} refers to when they convey the same meaning, regardless of structure. The choice between syntactic and semantic equivalence depends on the task. For instance, NLI with its triary output would use syntactic equivalence (see Figure~\ref{fig:example}), while free-form QA would use semantic equivalence (often implemented with BERT-like word embeddings \cite{mr-mtqas, mr-nattestgen}). Similarly, the second most common output relation is \texttt{Difference}, and follows the same principle. 



\smallskip
Many MRs share commonalities, but are distinct enough to be considered separate.  One example is the varying levels of granularity that different papers use in defining MRs. For instance, MR-8~\cite{mr-metal} involves creating semantically similar sentences by substituting
words with their synonyms. MR-17~\cite{mr-chinesellm} does this also, but only substitutes nouns. MR-72~\cite{mr-evalnli} adapts this specifically for NLI, substituting words only in the \texttt{premise} of the input. These variations may have different fault detection capabilities, and so are kept separate.

\smallskip
\noindent
\textbf{Task discussion:}
We identified 24 distinct NLP tasks, described in Table~\ref{tab:task}. It is important to mention that while some MRs are specific to certain tasks (e.g., MR-188 is specific to Translation (TR)), others have been applied to multiple tasks. For example, we found that MR-1 to MR-20 were applied across various tasks, with MR-1 to MR-4 being applied to the highest number (SA, TC, TD, QA, SM, IR) (see \textit{footnote$^a$} of Table~\ref{tab:mr_table} for the description of grouped notation $G_x$). Moreover, Table~\ref{tab:mr_table} associates tasks to MRs based on their use in the papers we reviewed, not necessarily the tasks to which they could be applied in other contexts. Indeed, MR-1 to MR-4 are presumably general enough to be applied to any NLP task.

\smallskip
To note, the number of MRs for a task (Table~\ref{tab:task}) is not an indication of how well-studied that task is with MT. For instance, sentiment analysis (SA)---with the highest number of MRs---derives 20 out of its 32 relations from a single paper (Table~\ref{tab:mr_table}), the rest from only three others. In contrast, Translation (TR) sources its 24 MRs from 14 different studies.







\section{Experiments}
We conducted a series of experiments to address three research questions:

\vspace{2mm}
\begin{itemize}
    \item[\textbf{RQ1}] \textbf{Failure Rate.} \textit{To what extent does MT exhibit failures in LLMs?}
    \item[\textbf{RQ2}] \textbf{Comparison.} \textit{How does MT compare with traditional testing that uses labeled data?}
    \item[\textbf{RQ3}] \textbf{Manual Validation.} \textit{What is the true positive rate of the reported failures?}
\end{itemize}

\vspace{2mm}

\textbf{RQ1} evaluates MT's effectiveness at detecting oracle violations by running \numtests test groups (source and follow-up pairs) from \nummr MRs on three popular LLMs. \textbf{RQ2} compares the results with the source input's ground truth provided by the datasets. Although MT is typically not applied when oracles (e.g., labeled data) are present, we used labeled data to better understand MT's fault detection capabilities. \textbf{RQ3} assesses MT's reliability through manual analysis of detected failures, identifying true positives and limitations.

\smallskip
\noindent
\textbf{Chosen tasks and relations:}
We selected four tasks from Table~\ref{tab:task}---question answering with context (QAc), natural language inference (NLI), sentiment analysis (SA), and relation extraction (RE)---as they provide a good range to evaluate LLMs' language understanding.

\smallskip
In Table~\ref{tab:mr_table}, we categorized the MRs based on the tasks they were proposed for. However, this does not mean these MRs cannot be applied to other tasks; indeed, several of the relations were presented for multiple tasks or explicitly stated to be generally applicable. Consequently, we chose \textbf{\nummr~metamorphic relations to implement}, spread across the four selected tasks, including MRs not necessarily originally proposed for that task. 
We selected these based on: (i) inclusion in METAL~\cite{mr-metal} (MRs 1–12), (ii) common use across tasks, (iii) MRs proposed for one task but suited to many, (iv) task-specific MRs, and (v) ease of understanding to reduce misimplementation. 
The first five columns of Table~\ref{tab:rq1-mr} show the implemented MRs and the tasks to which we applied them.

\smallskip
\noindent
\textbf{\tool implementation:}
To implement the input transformations $\rightsquigarrow$ of the \nummr MRs, we use two methods: function-based, and LLM-based. For simple procedures, we use traditional function-based methods to generate follow-up inputs (for MRs 1-7, 9, 19, 49, 84, 102, 120, 126, and 128).
For instance, randomly adding a sentence to the end of the input (MR-84) was done through concatenation, obtaining the sentences from an online source\footnote{https://randomwordgenerator.com/sentence.php}. As another instance, introducing keyboard mistakes (MR-126) was done through the NLPAug library\footnote{https://github.com/makcedward/nlpaug}.

For more complex cases, we utilise an LLM (\textsc{\hermes}). This is done via few-shot prompting. For example, we used the prompt 
\texttt{Paraphrase the following text: "\{TEXT\}" Only output the changed text, nothing else.} for the MR in Fig.~\ref{fig:example}. 
LLMs, unlike traditional techniques, consistently outputs text that is grammatically correct and thus viable test cases~\cite{natural-adversarial-examples-llm}. Without using LLMs, one would need check for grammatical consistency and text sensicality~\cite{aeon2022}. 

While we did not formally validate \hermes for input transformation, preliminary tests showed reasonable alignment with the MRs. \gpt may perform better, but \hermes was more practical due to \gpt rate limits. 

\smallskip
We crafted the LLM prompts for both the tasks and input transformations through an iterative process, following prompt engineering principles~\cite{chen2023unleashing} and refining them until we felt them sufficiently effective. 
For the transformation prompts, as we aimed to compare MRs across multiple tasks, we designed the few-shot examples generically to be applicable across different tasks, keeping the prompts and examples the same for all tests. 
We used zero-shot 
for the implementation of tasks as we believe this would give more insight into the generalisability of the task performance, as well as minimising bias from any examples we were to choose.
While we did not rigorously evaluate the prompts, preliminary checks show that they perform reasonably well for the tasks at hand. 

\smallskip
Although there is the risk of nondeterminism in the transformations performed by \hermes (unfortunately not supporting a random seed), we did not deem this a critical issue. This is because: (i) the stochastic effect is diluted due to applying the same input transformation to a myriad of different inputs; (ii) in preliminary runs, we did not observe significant differences when repeatedly transforming the same input; and (iii) repeating the experiments would require excessive time and resources. 

\smallskip
Some tasks have multiple components to their inputs---for example, NLI has a \texttt{premise} and a \texttt{hypothesis} (see Figure~\ref{fig:example}). We observe that, for these tasks, oftentimes previous research would apply the same transformation to each component and regard them as separate MRs (for example, MR-71 to 73). This process can be applied to many of the \nummr MRs we selected. Thus, for our chosen multiple-component tasks (QA and NLI), we test MRs on every combination when possible. For instance, in QA, we may apply the input transformation on the first component, then the second, then both, resulting in three distinct  metamorphic test cases. This effectively increases the variations of those MRs for these tasks as well as the number of metamorphic test groups generated. 

\smallskip
To measure semantic equivalence for computing the output relation, we use BERT \cite{devlin-etal-2019-bert} to assess the similarity between two texts, a common practice in NLP-related MT studies~\cite{mr-mtqas, mr-nattestgen}. Specifically, we use \textsc{paraphrase-MiniLM-L6-v2}
, a popular model on HuggingFace. If the similarity score is above a threshold, we consider the texts to be semantically equivalent. To decrease potential false positives, we selected different thresholds for MRs with equivalence and difference relations. 
To obtain these thresholds, we initially adopted a threshold of 0.6 (inspired by METAL~\cite{mr-metal}, though they employed a different output comparison method) and applied it to preliminary test data consisting of metamorphic output pairs. Through manual analysis, we selected the thresholds where 75\% of the true positives were retained. 
This resulted in a similarity threshold of 0.8, and a dissimilarity threshold of 0.4. 


\begin{table}[t!]
\centering
\rowcolors{2}{gray!10}{white} 

\setlength{\tabcolsep}{10pt}
\renewcommand{\arraystretch}{0.9}
\caption{Datasets used in our experiments}
\label{tab:dataset_table}
\resizebox{\columnwidth}{!}{%
\begin{tabular}{llrrl}
\toprule
\textbf{NLP Task} & \textbf{Dataset Name} & \textbf{Avg. words} & \textbf{Size} & \textbf{Ref.} \\ \midrule
QA   & \textsc{Squad2}  & 137.0      & 142,000        & \cite{squad2-rajpurkar-etal-2018-know}    \\
NLI  & \textsc{SNLI}    & 20.4       & 570,000        & \cite{snli-young-etal-2014-image}    \\
SA   & \textsc{SST2}    & 19.2       & 70,000        & \cite{sst2-socher-etal-2013-recursive}    \\
RE   & \textsc{Re-DocRED}  & 182.4        & 4,050        & \cite{redocred-tan2022revisiting}    \\ \bottomrule
\end{tabular}
}
\end{table}
\smallskip
\noindent
\textbf{Datasets: }
For each task, we selected an existing labeled dataset that is popularly used in the reviewed papers. Table~\ref{tab:dataset_table} provides a summary of the four datasets. 
Due to the high computational cost of running experiments with LLMs and the large size of the datasets, we sampled them as described in Section~IV-E. The sampled instances serve as our source input texts for metamorphic testing. It is important to note that MT does not require ground truth (e.g., labeled data); however, we chose these datasets with labeled data to compare the effectiveness of traditional testing with MT (RQ2).

\smallskip
\textsc{SQuAD2}~\cite{squad2-rajpurkar-etal-2018-know} is a popular question answering dataset, with five of the six MT for free-form question answering (QA) papers using it \cite{mr-nattestgen, mr-qaasker1, mr-qaasker2, mr-qatest, mr-mtqas}. \textsc{SNLI}~\cite{snli-young-etal-2014-image} is a widely used NLI dataset from Stanford University and is used in the primary source of MRs for NLI~\cite{mr-evalnli}. \textsc{SST2}~\cite{sst2-socher-etal-2013-recursive} is a popular sentiment analysis dataset from Stanford University. \textsc{Re-DocRED}~\cite{redocred-tan2022revisiting} is a newer version of \textsc{DocRED}, a popular RE dataset. For each input, \textsc{Re-DocRED} provides several possible relations; we randomly selected one for each instance.

\smallskip
\noindent
\textbf{LLMs under test: }
We performed our experiments on three LLMs. These are chosen as they are state-of-the-art models, and are readily available at our Institution.

\begin{itemize}
\item \textbf{\texttt{gpt-4-1106}}\footnote{\url{https://platform.openai.com/docs/models}} (\textbf{\gpt})
from OpenAI.

\item \textbf{\texttt{llama-3.1-70b-instruct}}\footnote{\url{https://huggingface.co/meta-llama/Meta-Llama-3-70B-Instruct}} (\textbf{\llama})
from META, comparable to GPT-4 in many areas.

\item \textbf{\texttt{nous-hermes-2-mixtral-8x7b-dpo}}\footnote{\url{https://huggingface.co/NousResearch/Nous-Hermes-2-Mixtral-8x7B-DPO}} (\textbf{\hermes}),
an open-source model. A fine-tuned version of \textsc{Mixtral8x7b}.

\end{itemize}

\smallskip
\noindent
\textbf{Experimental setup:} Using our four chosen tasks and \nummr relations, \tool takes input data from a dataset and transforms each input into one or more follow-up inputs, forming so-called \textbf{metamorphic test groups}---groups composed of a source input and its corresponding follow-up(s). It then checks against the output relation to determine possible metamorphic oracle violations.  Throughout, it assesses whether all source/follow-up inputs/outputs adhere to any additional requirements of the MR. If they do not, that metamorphic group is discarded.
We run 1,000 instances randomly sampled from each dataset for each task.
This is performed for each MR
and corresponding task, 
resulting in 108 task-MR pairs for each LLM.
Through this, we get \numtestsunique unique metamorphic groups. We run these same groups for each LLM to obtain comparable results.
In total, discounting discarded groups, we run \numtests metamorphic groups, spanning four tasks, three LLMs, and \nummr MRs.

\subsection{RQ1: Failure Rate}

Table~\ref{tab:rq1-mr} shows the results for RQ1 and RQ2 by MRs, aggregating the outcomes for different tasks and LLMs.

\smallskip
Column ``\textit{\# groups}'' gives the number of metamorphic test groups executed for each MR (range: 2,165–24,217; average: 15,265; median: 20,120). Each metamorphic group $t = \langle x_1, .... x_n \rangle$ involves the inputs/outputs of two or more invocations of the LLM under test (source and follow-up inputs, see Sec.~II). The number of groups varies across MRs as some were applied to only one task, and others to more (see Column ``\textit{NLP Tasks}''). Additionally, some instances from the datasets did not satisfy the input relation and were thus discarded. 

\smallskip
Column ``$\lambda$'' gives the \textbf{failure rate} \(\lambda = \frac{\# \text{violations}}{\# \text{test groups}}\), the ratio of test groups that lead to metamorphic oracle violations (i.e., the input relation $\mathcal{R}_i$ is true, but the output relation $\mathcal{R}_o$ is false). 
The failure rates range from 0.00 (0\%) to 0.80 (80\%), with an average of 0.18 (18\%) and median of 0.15 (15\%).

\begin{table}[t]\centering
\caption{Results by Metamorphic Relations (RQ1 and RQ2)}
\vspace{-1mm}
\label{tab:rq1-mr}

\rowcolors{2}{gray!10}{white} 


\setlength{\tabcolsep}{6pt}
\renewcommand{\arraystretch}{0.8}

\resizebox{\linewidth}{!}{%

\begin{tabular}{r|cccc|rc|cccc}
\toprule
& \multicolumn{4}{|c}{\textbf{NLP Tasks}} & \multicolumn{2}{|c}{\textbf{RQ1 -- failure rate}} & \multicolumn{4}{|c}{\textbf{RQ2 -- labelled data}}\\ \midrule
\textbf{MR} & \textbf{QA} & \textbf{NLI} & \textbf{SA} & \textbf{RE} & \textbf{\# groups} & \(\lambda\) & \textbf{\ding{192}} & \textbf{\ding{193}} & \textbf{\ding{194}} & \textbf{\ding{195}} \\
\midrule

1 & $\bullet$ & $\bullet$ & $\bullet$ & $\bullet$ & 23,288 & 0.23 & 0.53 & 0.23 & 0.15 & 0.08 \\
2 & $\bullet$ & $\bullet$ & $\bullet$ & $\bullet$ & 23,288 & 0.19 & 0.56 & 0.24 & 0.12 & 0.07 \\
3 & $\bullet$ & $\bullet$ & $\bullet$ & $\bullet$ & 23,256 & 0.11 & 0.63 & 0.27 & 0.06 & 0.05 \\
4 & $\bullet$ & $\bullet$ & $\bullet$ & $\bullet$ & 23,256 & 0.17 & 0.58 & 0.25 & 0.11 & 0.06 \\
5 & $\bullet$ & $\bullet$ & $\bullet$ & $\bullet$ & 23,256 & 0.12 & 0.62 & 0.26 & 0.07 & 0.05 \\
6 & $\bullet$ & $\bullet$ & $\bullet$ & $\bullet$ & 23,288 & 0.18 & 0.57 & 0.25 & 0.12 & 0.07 \\
7 & $\bullet$ & $\bullet$ & $\bullet$ & $\bullet$ & 23,292 & 0.15 & 0.60 & 0.26 & 0.09 & 0.06 \\
8 & $\bullet$ & $\bullet$ & $\bullet$ & $\bullet$ & 24,175 & 0.19 & 0.56 & 0.25 & 0.11 & 0.08 \\
9 & $\bullet$ & $\bullet$ & $\bullet$ & $\bullet$ & 11,216 & 0.17 & 0.55 & 0.29 & 0.09 & 0.07 \\
10 & $\bullet$ & $\bullet$ & $\bullet$ & $\bullet$ & 18,131 & 0.27 & 0.41 & 0.23 & 0.22 & 0.14 \\
19 & $\bullet$ &  &  & $\bullet$ & 5,178 & 0.14 & 0.46 & 0.40 & 0.03 & 0.11 \\
25 & $\bullet$ & $\bullet$ & $\bullet$ & $\bullet$ & 23,284 & 0.21 & 0.50 & 0.22 & 0.19 & 0.09 \\
34 & $\bullet$ & $\bullet$ & $\bullet$ & $\bullet$ & 23,284 & 0.39 & 0.34 & 0.14 & 0.35 & 0.17 \\
49 & $\bullet$ & $\bullet$ & $\bullet$ & $\bullet$ & 11,206 & 0.10 & 0.59 & 0.31 & 0.04 & 0.05 \\
51 & $\bullet$ & $\bullet$ & $\bullet$ & $\bullet$ & 23,262 & 0.17 & 0.59 & 0.24 & 0.10 & 0.07 \\
57 &  &  &  & $\bullet$ & 2,165 & 0.36 & 0.08 & 0.56 & 0.05 & 0.31 \\
77 &  & $\bullet$ &  &  & 3,013 & 0.02 & 0.70 & 0.26 & 0.02 & 0.02 \\
78 &  & $\bullet$ &  &  & 3,013 & 0.01 & 0.71 & 0.27 & 0.02 & 0.01 \\
79 &  & $\bullet$ &  &  & 3,013 & 0.00 & 0.72 & 0.27 & 0.01 & 0.00 \\
80 &  & $\bullet$ &  &  & 3,013 & 0.00 & 0.72 & 0.27 & 0.00 & 0.00 \\
84 & $\bullet$ & $\bullet$ &  & $\bullet$ & 14,217 & 0.11 & 0.60 & 0.29 & 0.05 & 0.06 \\
102 & $\bullet$ & $\bullet$ & $\bullet$ & $\bullet$ & 23,256 & 0.10 & 0.63 & 0.27 & 0.05 & 0.05 \\
120 & $\bullet$ & $\bullet$ & $\bullet$ & $\bullet$ & 23,262 & 0.16 & 0.59 & 0.25 & 0.09 & 0.06 \\
126 & $\bullet$ & $\bullet$ & $\bullet$ & $\bullet$ & 24,211 & 0.21 & 0.53 & 0.26 & 0.13 & 0.07 \\
127 & $\bullet$ & $\bullet$ & $\bullet$ & $\bullet$ & 24,217 & 0.20 & 0.54 & 0.26 & 0.13 & 0.07 \\
128 & $\bullet$ & $\bullet$ & $\bullet$ & $\bullet$ & 24,211 & 0.15 & 0.58 & 0.27 & 0.09 & 0.06 \\
136 & $\bullet$ & $\bullet$ &  & $\bullet$ & 20,243 & 0.15 & 0.58 & 0.27 & 0.09 & 0.06 \\
137 & $\bullet$ & $\bullet$ & $\bullet$ & $\bullet$ & 11,216 & 0.37 & 0.43 & 0.20 & 0.21 & 0.16 \\
141 &  &  &  & $\bullet$ & 2,165 & 0.06 & 0.05 & 0.47 & 0.07 & 0.42 \\
142 &  &  &  & $\bullet$ & 2,165 & 0.80 & 0.01 & 0.05 & 0.13 & 0.82 \\
149 & $\bullet$ &  & $\bullet$ & $\bullet$ & 14,223 & 0.10 & 0.62 & 0.28 & 0.05 & 0.05 \\
150 &  &  & $\bullet$ &  & 3,013 & 0.13 & 0.72 & 0.13 & 0.12 & 0.03 \\
151 &  &  & $\bullet$ &  & 3,013 & 0.14 & 0.70 & 0.13 & 0.14 & 0.02 \\
152 & $\bullet$ & $\bullet$ & $\bullet$ & $\bullet$ & 19,997 & 0.27 & 0.44 & 0.20 & 0.20 & 0.17 \\
154 &  &  & $\bullet$ &  & 3,013 & 0.21 & 0.64 & 0.11 & 0.20 & 0.05 \\
155 & $\bullet$ & $\bullet$ &  & $\bullet$ & 20,249 & 0.10 & 0.62 & 0.28 & 0.05 & 0.05 \\

\hiderowcolors \midrule
\textbf{AVG} &  &  &  &  & 15,265 & 0.18 & 0.54 & 0.26 & 0.11 & 0.10 \\
\textbf{MED} &  &  &  &  & 20,120 & 0.15 & 0.58 & 0.26 & 0.09 & 0.06 \\
\midrule
\textbf{TOT} &  &  &  &  & 549,548 & 0.18 
& & & & \\

\bottomrule

\end{tabular}
}
\end{table}

\smallskip
Tables~\ref{tab:rq1-tasks} and~\ref{tab:rq1-llm} summarize results by tasks and LLMs, respectively. QA tasks have the fewest failures ($\lambda=0.12$), while RE tasks have the most ($\lambda=0.32$). Among LLMs, \gpt has the lowest failure rate ($\lambda=0.14$). 

\begin{custombox}{Answering RQ1}
The failure rate $\lambda$ of the \nummr MRs averages to 18\%. There is significant variability in the ratio across MRs ($\lambda$ from 0\% to 80\%), even within the same task. 
\end{custombox}

\subsection{RQ2: Comparison with Traditional Testing}

To compare MT with traditional testing in the context of LLMs, we computed the confusion matrix of two oracles: the \textbf{metamorphic oracle}, and the \textbf{ground truth of the source output} as provided by the original labeled dataset. Table~\ref{tab:confusion} describes the notation for each of the four possible outcomes.

\begin{table}[t]
\caption{Results by NLP Tasks (RQ1 and RQ2)}
\vspace{-1mm}
\label{tab:rq1-tasks}
\rowcolors{2}{gray!10}{white} 


\setlength{\tabcolsep}{8pt}
\renewcommand{\arraystretch}{0.7}
\resizebox{\columnwidth}{!}{%

\begin{tabular}{l|rc|cccc}
\toprule
& \multicolumn{2}{|c}{\textbf{RQ1 -- failure rate}} & \multicolumn{4}{|c}{\textbf{RQ2 -- labelled data}} \\ \midrule

\textbf{Task} & \textbf{\# tests} & \(\lambda\) & \ding{192} & \ding{193}& \ding{194} & \ding{195} \\ \midrule

NLI & 200,987 & 0.17 & 0.59 & 0.23 & 0.13 & 0.06 \\
QA & 205,179 & 0.12 & 0.67 & 0.21 & 0.09 & 0.03 \\
RE & 67,958 & 0.32 & 0.09 & 0.57 & 0.04 & 0.30 \\
SA & 75,424 & 0.25 & 0.59 & 0.13 & 0.22 & 0.05 \\

\midrule \hiderowcolors

\textbf{AVG} & 137,387 & 0.22 & 0.48 & 0.29 & 0.12 & 0.11 \\
\textbf{MED} & 138,205.5 & 0.21 & 0.59 & 0.22 & 0.11 & 0.05 \\
\bottomrule

\end{tabular}
}
\end{table} 
\begin{table}[t]\centering
\vspace{-3mm}
\caption{Results by Large Language Models (RQ1 and RQ2)}
\label{tab:rq1-llm}
\vspace{-1mm}
\rowcolors{2}{gray!10}{white} 


\setlength{\tabcolsep}{6.5pt}
\renewcommand{\arraystretch}{0.7}

\resizebox{\linewidth}{!}{%

\begin{tabular}{l|rc|cccc}
\toprule
& \multicolumn{2}{|c}{\textbf{RQ1 -- failure rate}} & \multicolumn{4}{|c}{\textbf{RQ2 -- labelled data}} \\ \midrule

\textbf{LLM} & \textbf{\# tests} & \(\lambda\) & \ding{192} & \ding{193}& \ding{194} & \ding{195} \\ \midrule

\gpt & 169,787 & 0.14 & 0.64 & 0.21 & 0.10 & 0.05 \\
\llama & 191,082 & 0.18 & 0.53 & 0.28 & 0.11 & 0.08 \\
\hermes & 188,679 & 0.21 & 0.51 & 0.26 & 0.13 & 0.10 \\

\midrule \hiderowcolors

\textbf{AVG} & 183,183 & 0.18 & 0.56 & 0.25 & 0.11 & 0.08 \\
\textbf{MED} & 188,679 & 0.18 & 0.53 & 0.26 & 0.11 & 0.08 \\

\bottomrule

\end{tabular}
}
\end{table}

\begin{figure*}[ht]
\centering
  \vspace{-1mm}
  \includegraphics[width=0.95\linewidth]{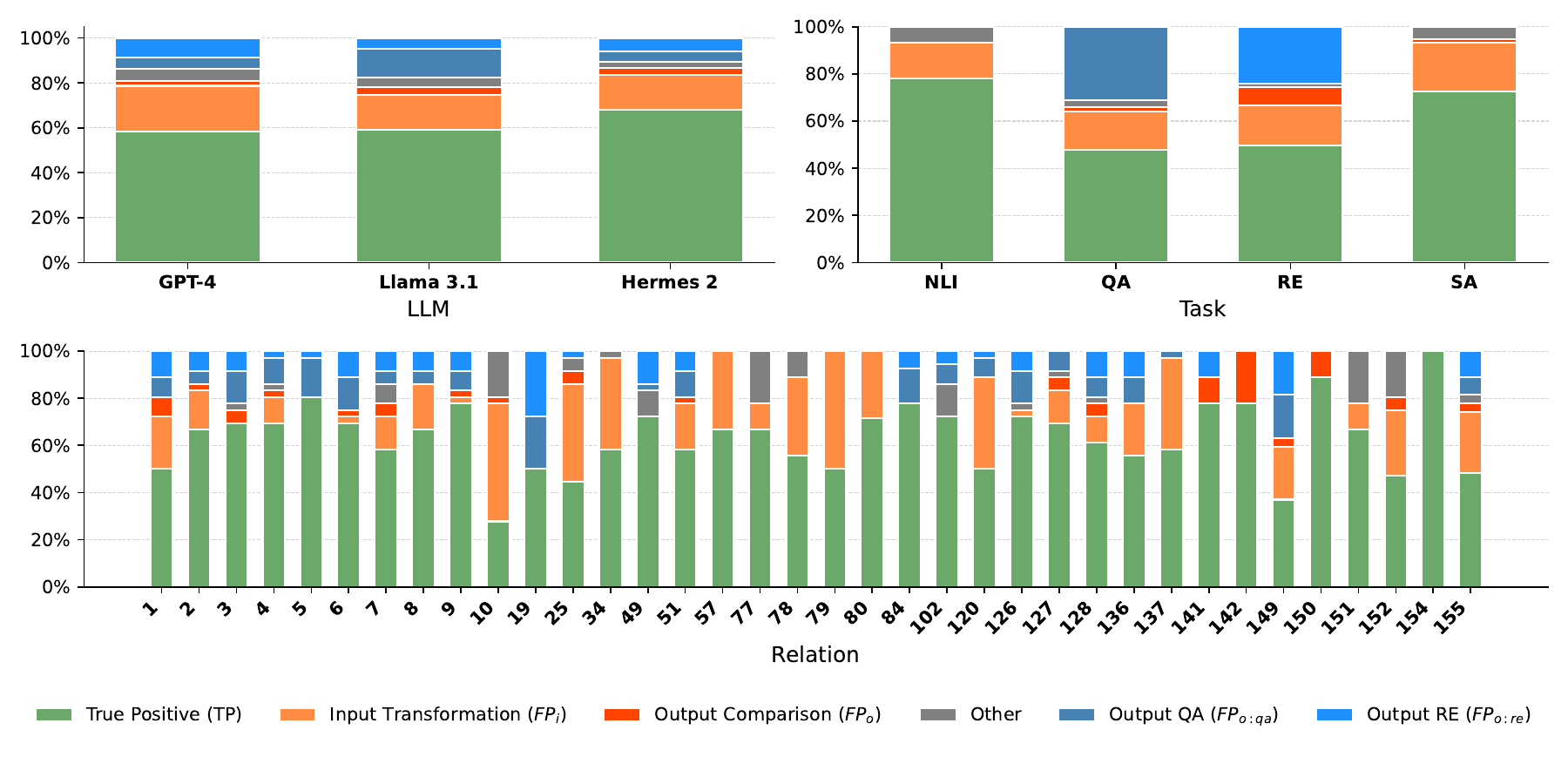}
  \vspace{-8mm}
  \caption{Stacked bar chart of the manual labels by  LLM, Task, and MR (RQ3)}
  \vspace{-3mm}
  \label{fig:label-mr}
\end{figure*}

\smallskip
Following Section~II, consider a metamorphic test group \( t = \langle x_1, \ldots, x_n \rangle \) 
with an associated MR \( \mathcal{R}_i (x_1, \cdots, x_n) \Rightarrow \mathcal{R}_o(f(x_1), \cdots, f(x_n)) \).
The metamorphic test oracle fails (an oracle violation indicating incorrect behavior) if this 
Boolean implication is false
(\ding{194} or \ding{195}). Otherwise, it passes (\ding{192} or \ding{193}).

\smallskip
For the same group \(t\), let \(\hat{f}(x_1)\) be the ground truth (from a labeled dataset) of the expected output 
for the source input \(x_1\). The ground truth oracle fails if 
\(\hat{f}(x_1) \neq f(x_1)\) (\ding{193} or \ding{195}), and passes otherwise (\ding{192} or \ding{194}). 
When tasks require semantic equivalence 
for the comparison,
we follow the method described in Section~IV; otherwise, we compare the strings syntactically
(e.g., \texttt{neutral} versus \texttt{entailment} in Figure~\ref{fig:example}).

\begin{table}[t!]

\caption{Confusion matrix of the MR and ground-truth oracles (RQ2)}
\vspace{-1mm}
\setlength{\tabcolsep}{2pt}
\renewcommand{\arraystretch}{1}
\resizebox{\columnwidth}{!}{%
\begin{tabular}{c|cc|llrl}
\toprule
& \multicolumn{2}{c|}{$\mathbf{\hat{f}(x_1) \neq f(x_1)}$} & & \ding{192} & $\sim53\%$ &Both oracles are passing \\ 
\textbf{MR(t)} & \textbf{True} & \textbf{False}& & \ding{193} & $\sim27\%$ & MT oracle passes, source output is incorrect \\  \cmidrule{2-3}
\textbf{True} & \ding{192} & \ding{193} & & \ding{194} &  $\sim11\%$ & MT oracle fails, source output is correct\\
\textbf{False }& \ding{194} & \ding{195} & & \ding{195} & $\sim10\%$ & Both oracles are failing \\
\bottomrule
\end{tabular}
}
\label{tab:confusion}
\end{table}

\smallskip
It is important to clarify that these oracle results are not directly comparable. On one hand, we only have the ground truth for the expected output of the source input, not for the metamorphic relation or any follow-up outputs. On the other hand, a failure in the metamorphic oracle does not necessarily indicate that the source output is incorrect; it may also point to issues with the follow-up output(s), or both. Nevertheless, we can still draw meaningful conclusions by analysing the results.

\smallskip
We partition the \numtests test groups into four disjoint sets (\ding{192}, \ding{193}, \ding{194}, \ding{195}) depending on whether the MR was satisfied or violated, and whether the source output was correct or not according to the ground truth. Columns \ding{192}, \ding{193}, \ding{194}, and \ding{195} in Tables~\ref{tab:rq1-mr}, \ref{tab:rq1-tasks}, and \ref{tab:rq1-llm} present the confusion matrix results grouped by MRs, tasks, and LLMs, respectively. Interestingly, the average ratios across the three are quite similar (with some exceptions). Thus, we can make some common observations:

\medskip
\textbf{\ding{192}} ($\sim$53\% of all groups) Both oracles pass. This is the most common case, which makes sense as the tested LLMs generally have high capabilities, leading to mostly correct executions.

\textbf{\ding{193}} ($\sim$27\% of all groups) The metamorphic oracle fails to detect that the source output is incorrect. An incorrect source output often leads to similarly incorrect follow-up outputs, likely making metamorphic fault detection more challenging.

\textbf{\ding{194}} ($\sim$11\% of all groups) The metamorphic oracle detects a faulty behavior that the ground truth oracle does not.
As the ground truth oracle on the source output passes,
there is likely a problem in the follow-up output. 
This is the most interesting case, showing the complementary nature of the two oracles.

\textbf{\ding{195}} ($\sim$10\% of all groups) Both oracles fail, indicating overlapping fault detection capabilities.

\medskip

The exceptions to this trend are MR-57, 141, and 142---all applied to RE---and the overall RE task itself. They show fewer MR satisfactions (\ding{192}) and more violations (\ding{194}). This likely stems from RE allowing multiple correct answers (e.g., ``David Bowie" and ``Space Oddity" could relate as both ``singer" and ``author"). As a result, our semantic similarity method may misclassify valid variations as failures.


\begin{custombox}{Answering RQ2}
Traditional label-based testing detects more faults overall, but MT complements it by identifying faults missed by labeled data, at zero human cost. Thus, MT is valuable when labeled data is limited or costly.
\end{custombox}

\subsection{RQ3 - Manual Validation}\label{subsec:RQ3 - Manual Validation}

To validate the results, we manually assessed a sample of the metamorphic violations to determine whether they were true positive faulty behaviors. 
To reduce bias toward high-violation MRs, we equally sampled from all combinations of LLM, task, and MR. We randomly selected three for each combination, totaling 967 violations distributed among three authors. 


\begin{table}[t]
\vspace{-4mm}
\caption{Classification of Manual Validation Results (RQ3)}
\vspace{-1mm}
\label{tab:labels}
\rowcolors{2}{gray!10}{white} 


\setlength{\tabcolsep}{2pt}
\renewcommand{\arraystretch}{0.9}
\resizebox{\columnwidth}{!}{%
    \begin{tabular}{lp{0.9\linewidth}}
        \toprule
        \textbf{Label} & \textbf{Description - True Positive (TP) and False Positive (FP)} \\
        \midrule
        \textbf{TP} & The metamorphic relation correctly identified a faulty behavior according to human judgment. \\
        \textbf{FP$_\text{i}$} & The MR failed because the input transformation altered the source input either too much or too little, depending on whether equivalence or difference was intended. \\
        \textbf{FP$_\text{o}$} & The output relation failed to correctly compare outputs (e.g., BERT's semantic similarity misidentifying equivalence or difference). \\
        \textbf{FP$_{\text{o:qa}}$} & In QA tasks, outputs indicate the answer is unknown, but the output comparison fails to recognize equivalence. \\
        \textbf{FP$_{\text{o:re}}$} & In RE tasks, outputs specify different but correct textual relations, and the output comparison fails to recognize correctness. \\
        \textbf{FP$_{\text{MR}}$} & The input and output relations are correct, but the MR is nonsensical for the given inputs. \\
        \textbf{FP$_{\text{other}}$} & Uncategorized cases (e.g., empty LLM output). \\
        \bottomrule
    \end{tabular}
}
\end{table}

\smallskip
Each evaluator examined the inputs, outputs, and MR descriptions. To facilitate this, we built a web interface highlighting the textual difference between source and follow-up inputs/outputs. The evaluators chose from the seven labels described in Table~\ref{tab:labels}.
Note that FP$_\text{o}$, FP$_\text{o:qa}$, and FP$_\text{o:re}$ are disjoint sets. We chose to label FP$_\text{o:qa}$ and FP$_\text{o:re}$ separately from other output relation issues
due to their prominence.
\smallskip

\smallskip
Figure~\ref{fig:label-mr} shows the stacked bar charts of the results grouped by MR, task, and LLM, respectively. As only one violation (of MR-155) was deemed nonsensical,
we incorporated FP$_\text{MR}$ into FP$_\text{other}$. On average, the TP rate is $62\%$ across all 967 metamorphic oracle violations manually analysed. While the bar chart of LLMs shows a similar proportion across different LLMs (left top corner), the chart divided by task (right top corner) provides more insight into which tasks MT is more prone to have false positives. As expected, QA and RE pose the greatest challenges because their free-form outputs require  semantic comparisons. The NLI task had no issues with the output relation as it uses syntactic equivalence among three possible outputs (see Figure~\ref{fig:example}). Similarly, sentiment analysis had fewer issues due to the ease of comparison of its numerical sentiment score. False positives derived from input transformation issues is the most common category.

\begin{custombox}{Answering RQ3}
True positives are the most common ($62\%$), but some MRs have significant false positives.
\end{custombox}

\section{Discussion and Analysis}

\begin{table}[t]
\caption{Reported False Positive rates of previous work on MT for NLP}
\label{tab:compare-precision}
\rowcolors{2}{gray!10}{white} 
\centering
\setlength{\tabcolsep}{3pt}
\renewcommand{\arraystretch}{0.9}
\begin{tabular}{@{}llll@{}}
\toprule
\textbf{Task} & \textbf{Ref.}       & \textbf{Lower} & \textbf{Upper} \\ \midrule
TR & \cite{mr-treeprune}   & 0.16 & 0.38 \\
TR & \cite{mr-wordclosure} & 0.30 & 0.43 \\
TR & \cite{mr-constinv}    & 0.13 & 0.37  \\
TR & \cite{mr-reftrans}    & 0.00 & 0.22     \\
TR & \cite{mr-patinv}      & 0.23 & 0.57 \\
TR & \cite{mr-structinv}   & 0.22 & 0.30  \\
TR & \cite{mr-automt}      & 0.09 & 0.23  \\
TR & \cite{mr-backtrans}   & 0.22 & 0.30 \\ \bottomrule  \hiderowcolors
\end{tabular}%
\quad
\begin{tabular}{@{}llll@{}}

\toprule
\textbf{Task} & \textbf{Ref. }      & \textbf{Lower} & \textbf{Upper} \\ \midrule \showrowcolors
QAb & \cite{mr-readcomp}     & 0.07  & 0.07  \\
QA & \cite{mr-qaasker1}      & 0.00  & 0.19     \\
QA & \cite{mr-qaasker2}      & 0.00  & 0.38     \\
QAc & \cite{mr-nattestgen}   & 0.02  & 0.03  \\
NER & \cite{mr-autoner}      & 0.02  & 0.20  \\
CR & \cite{mr-coref}         & 0.00  & 0.00     \\
Various & \cite{mr-intergen} & 0.02  & 0.08 \\  &  &  & \\ \bottomrule

\end{tabular}%
\end{table}

\textbf{False positive discussion: }
We observe a high false positive rate in our results. This is somewhat expected, as it is an issue inherent to MT for NLP. In traditional MT for software systems, MRs are usually defined exactly, such that the description and the implementation are one and the same~\cite{2017-chen-cs}. Under such conditions, if a false positive occurs, the MR itself must have been incorrectly specified and thus not a necessary property of the system under test~\cite{1998-chen-tr}. For example, consider the MR $x^2 = (-x)^2$ for the square function.
Given a number $x$, we can apply the input transformation precisely to obtain $-x$. The output relation of numerical equivalence ($=$) admits no FPs.

\smallskip
Conversely, in NLP, the MR serves more as a guideline; there may be many possible implementations of the same MR. 
Consider MR-35. The input relation \emph{``Replace keywords with unrelated terms''} specifies a vague implementation of detecting keywords and identifying unrelated terms, a task complicated by the inherently ambiguous 
nature of language. The semantic nature of the output relation further increases FPs.

\smallskip
Table~\ref{tab:compare-precision} summarizes the false positive rates reported by previous work on MT for NLP, showing both the lower and upper bounds from their experiments. Our results are comparable to these. This suggests that MT for LLMs does not produce substantially more false positives than traditional MT for NLP, which is an important finding.  Notably, prior MT research on LLMs~\cite{mr-metal} did not evaluate FPs.

\smallskip
We find that the most common cause FPs is input transformation errors. These issues typically fall into one of two categories: (i) the transformation changes too little even though the MR expects a different outcome, or (ii) it changes too much while the MR still expects an equivalent result (e.g. MR-127 misspell ``What is the \textbf{capital} of Chile?" $\rightsquigarrow$ ``What is the \textbf{apitpul} of Chile?").  Regarding output relation FPs, 
the BERT-based semantic comparison commonly used in NLP cannot as easily be used in LLM systems. 
In the QA task, $31\%$ of the false positives was due to the cosine similarity not being able to recognise the equivalence between, for instance, ``unknown" and a similar response.
In RE, $24\%$ of FPs occur because standard NLP assumes a single correct label, while LLMs can produce multiple valid answers (e.g., \emph{``son''} or \emph{``successor''} for \emph{``What is the relation between Charles I and Charles II?''}). These are intrinsic issues stemming from natural language's ambiguity and complexity.


\smallskip
While our results indicate that further research is needed to enhance the effectiveness of input transformations and output comparisons, it is unrealistic to expect these to perform flawlessly across all tasks and contexts.
A more attainable goal is to devise methods for assessing the confidence or quality of such transformations and comparisons,
thereby filtering out cases likely to be false positives.
For instance, prioritising failures by metrics such as MR type and the magnitude of similarity score would help reduce the cost of inspecting failing test cases in practical use.
In addition, if the oracle violations are used in an automated pipeline (e.g., for self-supervised fine-tuning or adversarial training), some FPs are less critical.

\smallskip
\textbf{Task dependency discussion:}
Our experiments confirm that some MRs exhibit varying TP rates depending on the task. For example, synonym substitution in QA (\emph{``What's the tallest building?"} $\rightsquigarrow$ \emph{``What's the highest building?"}, which remains equivalent) behaves differently in SA (\emph{``The movie was great"} $\rightsquigarrow$ \emph{``The movie was superb"}, which may increase the sentiment score). However, certain MRs (e.g., MR-9, MR-102)
prove effective across multiple tasks, consistently maintaining a high TP rate. Notably, we applied some MRs to tasks they were not originally designed for and found them to be effective. This is crucial, as it suggests that some MRs can be leveraged universally, demonstrating the potential of MT for evaluating locally deployed, fine-tuned LLMs.

\smallskip \textbf{Flakiness analysis:}
LLMs, compared to many NLP techniques, are non-deterministic in nature\footnote{Even when the temperature parameter is set to zero, some models or tokenizers may still randomly choose between tokens with equal scores.}. Therefore, it is important to understand how flaky (inconsistent) the results may be. We selected all 99,099 metamorphic groups that initially failed and re-ran them 9 additional times
to measure flakiness across 10 total runs. We used exactly the same groups for each run, thus guaranteeing no variability in generating the follow-up test cases and ensuring consistency across all experiments. 

\smallskip
Table~\ref{tab:flaky} shows the results (\textbf{all} row). 
Most metamorphic groups (62\%) consistently failed in the majority of runs (6-10 times), with 28\% failing in all 10 runs.
These results indicate that the observed issues reliably trigger MR violations and are not artifacts of randomness behaviors inherent to LLMs models. Moreover, although we re-ran only failing metamorphic groups rather than all \numtests groups (due to computational constraints), our results suggest that even a few runs may still reliably detect MR violations. 

\smallskip
Table~\ref{tab:flaky} also breaks down the failure rates between failure types. The average 10/10 rates between true (TP 21\%) and false (FP$_{all}$ 23\%) positives are similar, unexpectedly. 
This is likely coincidental, however, as the 10/10 rates for input-based false positives (46\%) differ greatly from that of output-based (0\%-11\%). The former is LLM-independent, leading to a higher chance of consistent violation; the latter largely depends on the LLM's nondeterministic output, often resulting in true satisfaction once the output has changed. 

\begin{table}[t]\centering
\caption{Flakiness of metamorphic groups with at least one failure}
\label{tab:flaky}

 \rowcolors{2}{gray!10}{white} 

\setlength{\tabcolsep}{3pt}
\renewcommand{\arraystretch}{0.8}

\resizebox{\linewidth}{!}{%

\begin{tabular}{l|rrrrrrrrrr}

\toprule
 & \multicolumn{10}{c}{\textbf{\# failures per 10 re-runs}} \\ 
\midrule
\textbf{Failure Type} & \textbf{1/10}  & \textbf{2/10 }  & \textbf{3/10}  & \textbf{4/10}  & \textbf{5/10}  & \textbf{6/10}  & \textbf{7/10}  & \textbf{8/10}  & \textbf{9/10}  & \textbf{10/10}  \\
\midrule
\textbf{all} & 0.08 & 0.07 & 0.08 & 0.08 & 0.08 & 0.08 & 0.08 & 0.08 & 0.10 & 0.28 \\
TP & 0.08 & 0.08 & 0.14 & 0.06 & 0.11 & 0.08 & 0.06 & 0.09 & 0.10 & 0.21 \\
FP$_\text{all}$ & 0.08 & 0.10 & 0.11 & 0.10 & 0.11 & 0.06 & 0.06 & 0.07 & 0.09 & 0.23 \\
FP$_\text{i}$ & 0.03 & 0.04 & 0.03 & 0.03 & 0.04 & 0.04 & 0.10 & 0.05 & 0.16 & 0.46 \\
FP$_\text{o}$ & 0.11 & 0.05 & 0.16 & 0.21 & 0.11 & 0.00 & 0.05 & 0.11 & 0.11 & 0.11 \\
FP$_\text{o:qa}$ & 0.22 & 0.18 & 0.22 & 0.13 & 0.13 & 0.09 & 0.02 & 0.02 & 0.00 & 0.00 \\
FP$_\text{o:re}$ & 0.00 & 0.19 & 0.06 & 0.19 & 0.25 & 0.06 & 0.06 & 0.16 & 0.03 & 0.00 \\
FP$_\text{other}$ & 0.00 & 0.05 & 0.16 & 0.05 & 0.16 & 0.11 & 0.00 & 0.11 & 0.05 & 0.32 \\

\bottomrule

\end{tabular}
}
\end{table} 
\smallskip
\textbf{Envisioned use of \tool:} 
\tool can be combined with traditional testing by using MT to automatically identify inputs where the LLM fails to satisfy an MR. Traditional testing can then be applied to these inputs to define human-specified oracles, enabling more targeted validation while reducing the high cost of human labelling. 

\smallskip
\tool could be integrated into existing continuous integration/continuous deployment pipelines for LLM-based services. In particular, in the regression testing scenario, \tool could automatically detect if changes in the training, prompts, or model architecture impacted the performance of the system. This could be fully automated by checking if the failure rate significantly increases after modifications.

\section{Threats to validity}

\mypar{Data leakage: } A possible threat is data leakage, as the data we used was likely trained on by the LLMs. Indeed, during our manual inspection, we found that the LLM was sometimes responding too well on very perturbed follow-up inputs. However, this is not a severe issue; it simply means that the LLM performs better than it should. We are still able to detect issues using MT.


\mypar{Implementation issues: } Another potential threat concerns our implementation of prompts, input transformations, and output comparison. To mitigate this, we followed established practices from related work and applied prompt engineering techniques to refine our prompts where needed. All experimental data are publicly available\footref{link:doi}, and we welcome external validation.

\mypar{Input transformation bias:} In our experiments, we relied on \hermes to perform complex input transformations. However, \hermes is also one of the LLM under test, so the input transformation and output generation was done by exactly the same MR. This might have introduced biases. Performing more experiments possibly using a dedicated LLM for the input transformation is an important future work, as well as evaluating \tool on more recent and advanced LLMs (e.g., OpenAI's \textsc{o3}, which has advanced reasoning).

\section{Related Work}

To the best of our knowledge, this is the 
first comprehensive study of MT for LLMs through NLP. We now discuss the most related work in MT surveys, LLMs for MT, and MT for LLMs.

\mypar{MT surveys: } Previous surveys on MT have not specifically targeted MRs in the NLP domain. \textsc{METwiki}\footnote{Unreachable website \url{http://metwiki.net/}}~\cite{metwikipaper}
is a repository for retrieving MRs, containing MRs extracted from the MT survey by Segura et al.~\cite{Segura2016}. They do not discuss MRs for NLP, however. Their survey was conducted in 2016, while our literature search indicates that the first papers on MT for NLP date to 2018. Indeed, Figure 5 of Segura et al.'s survey, the MT applications over the years~\cite{Segura2016}, does not include NLP (or a related application). Another survey on MT by Xie et al.~\cite{xie2011testing} (conducted in 2011) provides a broader overview of MT techniques for machine learning but does not mention MT for NLP, which again aligns with our findings. Moreover, both surveys discuss the properties of MT rather than specific MRs.

\mypar{LLMs for MT: } There is growing interest in leveraging LLMs for various aspects of automated software testing, such as generating test cases and oracles~\cite{schafer2023adaptive,yuan2023no,Ruberto2025FromImplemented,ravi2025llmloop}. In the context of MT, recent research has explored using LLMs for MT of software systems~\cite{wang2024software,siddiq2024using,alshahwan2024automated}, including generating MRs~\cite{shin2024towards,tsigkanos2023large,luu2023can,srinivas2023potential,xu2024ase-mradopt}. Similarly, \tool employs LLMs for implementing MT, specifically for transformations. 
However, our focus is on testing LLMs, whereas these work leverages LLM to target MT for code-based software systems.

\mypar{MT for LLMs: } Adversarial attacks on prompts aim to perturbing \textit{prompts} to evaluate the qualities of LLMs~\cite{perez2022ignore,li2023multistepjailbreakingprivacyattacks,zhu2024promptrobustevaluatingrobustnesslarge,shen2023chatgpttrustmeasuringcharacterizing}. Some of these adversarial attacks partially resemble MRs and MT~\cite{zhu2024promptrobustevaluatingrobustnesslarge}. However, these approaches focus on altering the prompt to make the LLM respond differently, including jailbreak techniques that assign specific roles to prompts to induce privacy and security leaks~\cite{li2023multistepjailbreakingprivacyattacks}. In contrast, we focus on MRs applied to inputs rather than prompts, which is an orthogonal problem. Indeed, the MRs discussed in this paper are generic to any NLP technique, whereas adversarial attacks on prompts are specific to particular LLMs, with prompt-based jailbreak methods working for some LLMs but not others~\cite{li2023multistepjailbreakingprivacyattacks}. 


\smallskip
Regarding MT for LLMs in the NLP domain, 
\textsc{METAL}~\cite{mr-metal} is a recent framework for MT of LLMs. While the experimental settings of \textsc{METAL} are similar to \tool, including investigating using LLMs to implement input transformations, the two works differ substantially:
\begin{inparaenum}[(i)]
    \item \textsc{METAL} experimented with 13 MRs, whereas we chose \nummr, which includes all relevant MRs in \textsc{METAL}\footnote{   
    \textsc{METAL} includes additional MRs targeting more general LLM qualities (e.g., efficiency, fairness), which are outside the scope of this work.}. 
    \item The number of metamorphic test cases generated in our experiments is orders of magnitude higher. 
    \item \textsc{METAL} did not manually investigate the false positive rate of metamorphic oracle violations. We found that this is crucial to truly understand the capabilities and effectiveness of MT for LLMs. 
    \item We compared the MT results with the ground truth of the source inputs to provide more insights into the fault detection capabilities of MT.
\end{inparaenum}

\smallskip
Since we performed the literature search on MT for LLMs via NLP (up to 30 June 2024), a few papers have been published in this field.
There have been studies on detecting 
fact-conflicting hallucinations in LLM using MT~\cite{wu2025hallu, yang2025hallu,li2024drowzee, li2025hallutemporal}, mainly in the areas of question answering~\cite{wu2025hallu, yang2025hallu}. In addition, there have been papers on LLM-based dialogue systems~\cite{guo2025mortar} and evaluating LLM in-context learning via sentiment analysis and question answering~\cite{racharak2025incontext}.
However, these studies investigate specific applications of MT for LLMs under specific tasks, domain, or contexts. Conversely, our work aim at examining MT for LLM more generally; indeed, we still exceed all of these works in the breadth of our study, considering both the number of MRs and the range of NLP tasks evaluated.

\section{Conclusion and Future Work}
This paper presented the most comprehensive study on applying Metamorphic Testing (MT) to Large Language Models (LLMs). We systematically reviewed the existing literature and compiled a list of \nummrtotal metamorphic relations for NLP. We then introduced a framework called \tool that implements \nummr of these MRs, executing 561,297 metamorphic test groups on three LLMs across four NLP tasks. Using this collection of MRs, we explored the interactions between MRs, tasks, 
and LLM response semantics.

\smallskip
As this framework uses the general NLP testing method of Metamorphic Testing, it can also be used on any NLP system, not just LLMs. While \tool could still rely on the power of LLMs for implementing input transformations, we can replace the model under test with any NLP tool.  In particular, our catalog of 191 MRs represents the largest knowledge base of MRs for NLP to date, which we believe is a useful contribution for the entire NLP community beyond LLMs.

\smallskip
While MT for LLMs via NLP is promising, it remains a relatively new research direction, and several challenges remain. 
Based on our findings, we highlight several opportunities for future research.

\smallskip
\tool currently implements \nummr out of the \nummrtotal collected MRs. While implementing all MRs is a substantial task, we hope that releasing the source code\footref{link:gh}  will encourage the research community to contribute to \tool.

\smallskip
We investigated the same MRs applied across several tasks and found that many exhibit task independence. With LLMs increasingly being fine-tuned for tasks, it is important to find more of such task-independent MRs to more effectively test these systems automatically.

\smallskip
We found that false positive metamorphic violations is still a major challenge in MT for LLMs.  Given the inherent ambiguity of language, completely eliminating FPs is difficult. Future research should focus on detecting and filtering FPs to improve the effectiveness of metamorphic testing for LLMs.







\clearpage

\bibliographystyle{IEEEtran}
\bibliography{bibliography}

\end{document}